\documentclass[aps,pra,twocolumn,superscriptaddress]{revtex4}
\usepackage{bbding}
\usepackage{mathrsfs}
\usepackage{amsfonts}
\usepackage{amssymb}
\usepackage{color}
\usepackage{amsmath}
\usepackage{graphicx}
\usepackage{subfigure}
\usepackage{txfonts}
\usepackage[title]{appendix}
\usepackage{ulem}
\usepackage[colorlinks,linkcolor=blue,citecolor=blue]{hyperref}
\usepackage{mathrsfs}

\begin{document}

\title{Witnessing non-Markovianity with Gaussian quantum steering in collision model}

\author{Yan Li}
\affiliation{School of Physics, Dalian University of Technology, Dalian 116024, China}
\affiliation{Department of Physics, The Chinese University of Hong Kong, Shatin, New Territories, Hong Kong, China}

\author{Xingli Li}
\email{xinglili@cuhk.edu.hk}
\affiliation{Department of Physics, The Chinese University of Hong Kong, Shatin, New Territories, Hong Kong, China}

\author{Jiasen Jin}
\email{jsjin@dlut.edu.cn}
\affiliation{School of Physics, Dalian University of Technology, Dalian 116024, China}

\begin{abstract}
The nonincreasing feature of temporal quantum steering under a completely positive trace-preserving (CPTP) map, as proposed by Chen, {\it et al.} in Phys. Rev. Lett. 116, 020503 (2016), has been considered as a practical measure of non-Markovianity. In this paper, we utilize an all-optical scheme to simulate a non-Markovian collision model and to examine how Gaussian steering can be used as a tool for quantifying the non-Markovianity of a structured continuous variable (CV) Gaussian channel. By modifying the reflectivity of the beam splitters (BSs), we are able to tune the degree of non-Markovianity of the channel. After analyzing the non-Markovian degree of the dissipative channel within two steering scenarios, we discovered that the Gaussian steering-based non-Markovian measure depends the specific scenario because of the asymmetry of Gaussian steering. We also compared the Gaussian steering based non-Markovianity to the one based on the violation of the divisibility of CPTP map.
\end{abstract}
\date{\today}
\maketitle

\section{Introduction}
Einstein-Podolsky-Rosen (EPR) steering~\cite{EPR1935PR}, also known as quantum steering, is a phenomenon that can demonstrate non-local correlations between entangled systems. It refers to the ability of one observer, called {\it Alice}, can remotely manipulate the quantum state of another observer's, named {\it Bob}'s system by choosing which measurements to perform on her own part of an entangled system \cite{Uola2020RMP,Cavalcanti2017RPP,XiangPRXquantum2022}. This effect is a signature of entanglement and is related to Bell nonlocality, i.e., quantum steering is connected to entanglement and Bell nonlocality in that all steerable states are entangled, but not all entangled states are steerable; all states that violate the Bell inequalities, known as Bell nonlocal states, are also steerable, but not vice versa~\cite{Wiseman2007PRL}.

Quantum steering has been well-studied in the field of quantum mechanics for several decades since the concept was first introduced by Einstein, Podolsky, and Rosen in their pioneering paper \cite{EPR1935PR}. However, the topic has gained renewed interest in recent years, with numerous studies exploring its theoretical and practical applications in various areas  \cite{Wiseman2007PRL}. So far quantum steering has been studied more in-depth, with a focus on its potential applications in quantum communication and cryptography \cite{LiPRA2015,XiangPRAR2017,HePRA2018}, quantum metrology~\cite{YadinNC2021,LeePRR2023}, quantum networks~\cite{HuangPRA2019, LuPRL2020,JonesPRL2021}, information scrambling~\cite{LinPRA2021}, quantum computing~\cite{GheorghiuNJP2017} and the comprehensive reviews of the state of the art of quantum steering can be found in Ref.~\cite{Uola2020RMP,XiangPRXquantum2022}. Additionally, quantifying the non-Markovianity degree of the dynamics of quantum system has been identified as a potential application due to the concept of temporal steering proposed by Chen, {\it et al.} \cite{Chen2017PRL}, and the non-Markovian evolution of EPR steering has also been observed in recent experiments~\cite{WangPRL2023}.

Non-Markovian effect in the dissipative dynamics of an open quantum system refers to the situation that the evolution of  state of the system in the present time depends on its ``historic" states, which indicates the so-called memory effect~\cite{Angel2014Quantum,breuer2016colloquium,VegaRMP2017}.The non-Markovianity plays important roles in many areas of physics, including quantum information processing~\cite{AhnPRA2002, XuAoP2015,WhiteNC2020}, quantum optics~\cite{Crispin2004,TudelaQuantum2018}, and quantum thermodynamics~\cite{WhitneyPRB2021,DasPRA2021,CockrellPRE2022,ChoquehuancaPRA2023}. Thus, quantifying and measuring the non-Markovianity in dissipative quantum dynamics is always an active research area. Usually, the non-Markovianity of dissipative dynamics is quantified through two ways, one is introduced by Breuer-Laine-Piilo (BLP)~\cite{BreuerPRL2009}, which utilizes the distinguishability between two distinct initial states that undergo the same evolutionary process, the distinguishability of states is usually quantified by the trace distance. The other measure is proposed by Rivas-Huelga-Plenio (RHP)~\cite{RivasPRL2010}; the dynamic can be recognized as non-Markovian if its corresponding dynamical map $\Phi(t,t_{0})$ violates the CPTP-divisibility.  Besides, inspired by the RHP approach, Torre {\it et al}. have developed a measure of non-Markovianity for the Gaussian channel \cite{TorrePRL2015,TorrePRA2018}, which has provided a method without relying on the Choi-Jamiolkowski representation of the Gaussian channels~\cite{HolevoJMP2011} and without optimizing the set of input state~\cite{VasilePRA2011}.

Generally, the dissipative dynamics of an open quantum system can be described by quantum master equations~\cite{BreuerBook2002}. However, it is also worth mentioning that in recent years, the quantum collision model has become a powerful tool for constructing and investigating both the Markovian and non-Markovian dynamics in open quantum systems~\cite{GiovannettiPRL2021,CiccarelloPRA2013,JinPRA2015,CattaneoPRL2021,CiccarelloPR2022}. The reason behind the effectiveness of quantum collision models lies in their ability to discretize the infinite environmental degrees of freedom in an open quantum system as a family of identical states. This approach allows the model to express the interaction between the system and the environment as a series of ``collisions" between the system and the discrete environmental degrees of freedom. Thus, we can more concretely investigate the thermal and information exchange between the system and its surrounding environment~\cite{CiccarelloPR2022}. Additionally, after performing specific assumptions, this model is equivalent to the quantum master equation in the Lindblad form~\cite{CattaneoPRL2021}. As a result, the quantum collision models have been used not only to study non-Markovian dynamics and measure non-Markovianity~\cite{CamascaPRA2021,Li2022Entropy} but also to investigate quantum thermodynamics~\cite{ZhangPRA2021,ZhangPRA2023v1,ZhangPRA2023v2}, quantum synchronization~\cite{KarpatPRA2019,KarpatPRA2020,LiPRA2023}, and quantum many-body systems~\cite{ArisoyEntropy2019,LiPRA2020}.

Our research is focused on the time-evolution of Gaussian steering through the dissipative (Gaussian) channel. 
Unlike the previous studies~\cite{FrigerioPRA2021,SantisNJP2023}, here we employ the framework of the quantum collision models and propose an all-optical scheme that is possible to implement in experiment. Besides, compared with other measures, we find that only in the case of the auxiliary mode steering the system mode, the Gaussian steering based measure can faithfully reveal the non-Markovianity of the system's dynamics due to the asymmetric nature of quantum steering. We also notice that, during the evolution processes, Gaussian steering may show sudden death and birth behaviors, which have not been mentioned in previous works. Furthermore, within the quantum collision model in the continuous variable system that we are analyzing, we are able to initiate a ``dynamical transition" from Markovian to non-Markovian dynamics~\cite{KyawPRA2020}, or vice versa, by performing minor adjustments to a few parameters. In addition, this theoretical framework allows us to easily compare various non-Markovian measures.

This paper studies the possibility of witnessing and measuring non-Markovian dynamics via Gaussian steering in collision models. To this end, we introduce an ancilla mode and compare the non-Markovianity degrees under different Gaussian steering scenarios. We observe that due to the asymmetric nature of quantum steering, some parameter cases yield opposite dynamical properties in different scenarios. To avoid potential errors, we use another non-Markovian measure, the violation of CPTP map divisibility, and compare the results. We find that the non-Markovian region, as measured by the violation of CPTP map divisibility, aligns with the results obtained using the Gaussian steering, with the system being the steered component. We analyze the reasons for the differences between the two steering scenarios and explain the sudden changes
phenomena in the time-evolution process of Gaussian steering.

This paper is organized as follows. In Sec.~\ref{Sec:CollisionModel}, we introduce the basic concepts of our collision model and its corresponding all-optical scheme, then provide the characteristic function of the Gaussian state to explore the dynamical process. Then, in Sec.~\ref{Sec:NMmeasure}, the different non-Markovianity measures are introduced, including Gaussian steering and the violation of
CPTP map divisibility of the Gaussian channel. In Sec.~\ref{Sec:Result}, we present our main results, which include a comparison and discussion of the differences between the two steering scenarios. Additionally, we explain the factors that contribute to the occurrence of sudden changes in Gaussian steering. Finally, we summarize our findings in Sec.~\ref{Sec:Summary}.

\section{Simulation of Dissipative Dynamics using a Collision Model and its All-Optical Scheme}
\label{Sec:CollisionModel}
Deriving the non-Markovian quantum master equations can be a challenging task despite the availability of the efficient approaches such as the time-convolutionless (TCL) projection operator technique~\cite{ShibataJSP1977,ChaturvediZPB1979,UchiyamaPRE1999,BreuerAof2001}, the Nakajima-Zwanzig formalism~\cite{NakajimaPTP1958, ZwanzigJCP1960,VacchiniPRA2010,IvanovPRA2015}, etc. This is due to the inherent complexity of non-Markovian dynamics, which involve the so-called memory effects that are absent in Markovian cases~\cite{BreuerBook2002}. Therefore, constructing and measuring non-Markovian dynamics remain complex problems.

\begin{figure}[!htpb]
 \includegraphics[width=0.95\linewidth]{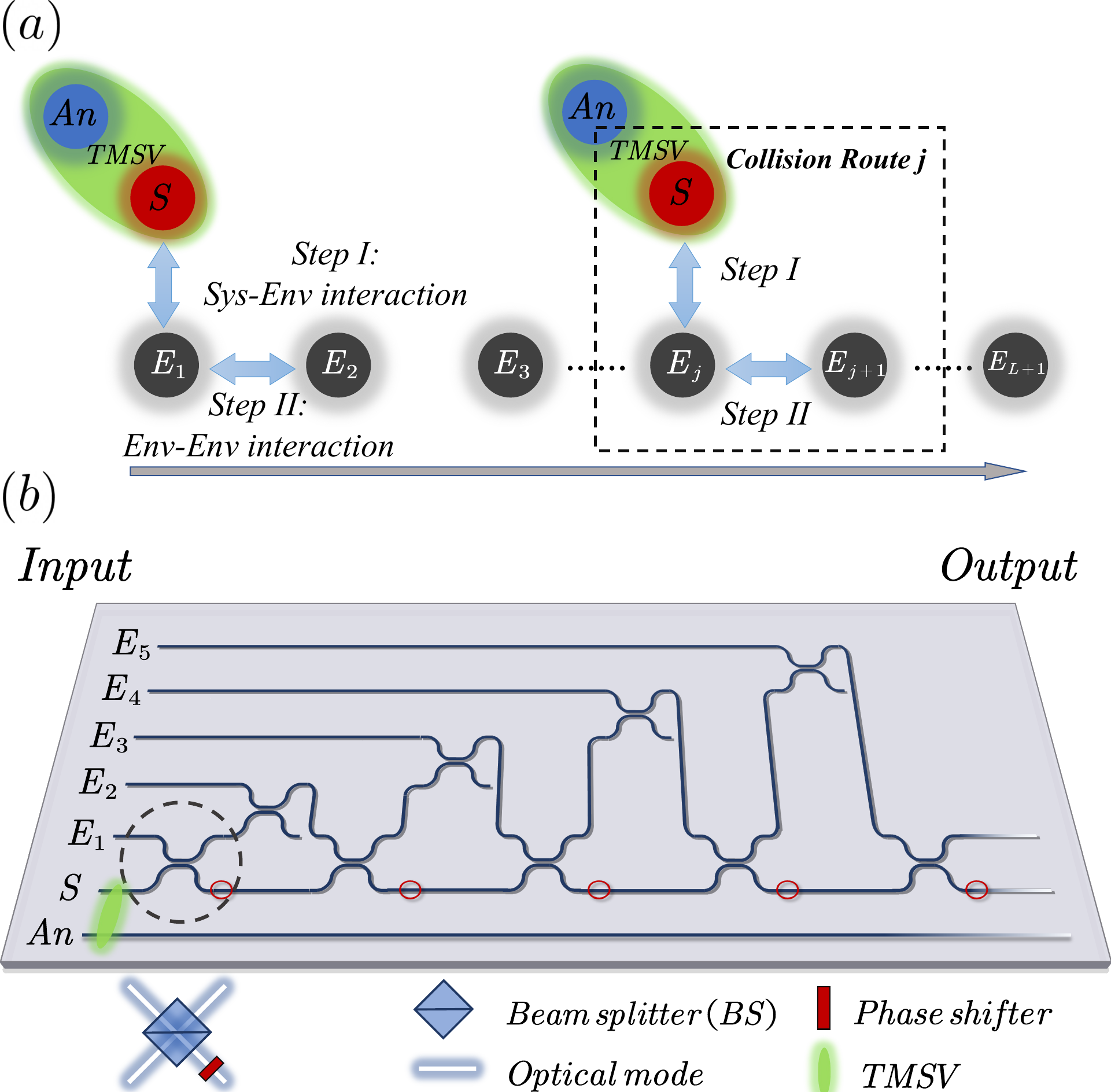}
  \caption{\label{FIG:Model} (a) Schematic diagram showing the block-interaction routes of collision model. The collision model consists of the ancilla $An$, the system $S$, and the environment blocks $E_{j}$, $j\in[1,L]$. The system is initialized to be entangled with the ancilla in the form of a two-mode squeezed vacuum (TMSV) state. Each collision route contains two separate steps in sequence: {\it Step I} describes the interaction that occurs between the system and the $j$-th environment block; {\it Step II} shows the internal interacting process of the environment blocks $E_{j}$ and $E_{j+1}$. (b) A pictorial illustration of the all-optical scheme for simulating the collisions is shown in (a), which contains BSs for introducing the mixing of optical modes. }
\end{figure}

Fortunately, after decades of development, collision models provide us with a new approach for investigating open quantum systems, especially the non-Markovian dynamics. In the collision model framework, the dissipative dynamics of a system can be described by its stroboscopic collisions with a collection of environmental ancillary quantum states. To be more specific, as shown in Fig.~\ref{FIG:Model}(a), the system $S$ is represented by a quantum state, while the environment is comprised of a series of identical states. Each block can be considered an environment state, and as the number of blocks increases, the degrees of freedom of the environment also increase, potentially meeting the requirement of infinite degrees of freedom. A collision is introduced by a composite unitary evolution between the different blocks, and by constructing various collision routes, the system exhibits a wide range of dynamic properties.

Our research, depicted in Fig.~\ref{FIG:Model}(a), concentrates on a collision scenario that is relatively simple yet facilitates the transition from Markovian to non-Markovian dynamics. A single collision cycle can be split into two sequential steps, exemplified by the $j$-th cycle: during {\it Step I}, system block $S$ collides with the $j$-th environmental block $E_{j}$, followed by an internal collision between environmental blocks during {\it Step II}, where $E_{j}$ interacts with $E_{j+1}$. It should be noted that if the system has no quantum correlations with block $E_{j}$ initially, during {\it Step I}, information from the system can flow into the environment, and in {\it Step II}, the information can flow into environmental block $E_{j+1}$ in the form of interactions. As a result, in the $j+1$-th collision cycle, the system will interact with the environmental block $E_{j+1}$, which contains its ``history". By adjusting the appropriate interaction strengths, the system's dynamics process can be made non-Markovian. In addition, to examine the dynamic properties of Gaussian steering and quantify non-Markovianity, we introduce an ancillary part and prepare it to be entangled with the system state. It is important to note that the ancilla does not evolve during the collision cycles.

Although constructing non-Markovian dynamics through a collision model is experimentally possible, precisely controlling the interaction strength and time between particles still remains challenging. However, the all-optical platforms own the advantages of high stability and arbitrary control of the device parameters and have already been used to experimentally realize Boson sampling~\cite{DengPRL2023,ZhongPRL2021} and quantum collision models~\cite{ChiuriSR2012,CuevasSR2019}. In this way, we propose an all-optical scheme as an ideal experimental platform for our research to improve the experimental implementation.

Our proposed all-optical scheme primarily involves an array of beam splitters, which we use to manipulate and mix different optical modes to realize ``interaction". Those independent optical modes propagating along different optical paths are regarded as the ``blocks" mentioned in the collision model. By constructing an appropriate optical path and BSs network, we can map the collision model into an all-optical system and obtain the corresponding output results based on the provided input patterns, as shown in Fig.\ref{FIG:Model}(b).

Here we recall the mathematical expression of mixing input optical modes $a^{\text{in}}_{1}$ and $a^{\text{in}}_{2}$ through BSs, which is in the following form,
\begin{eqnarray}
\begin{pmatrix}
 a_1^{\text{out}} \\[0.3em]
 a_2^{\text{out}} \\
\end{pmatrix}
=
\begin{pmatrix}
r&t\\
-t&r\\
\end{pmatrix}
\begin{pmatrix}
a_1^{\text{in}}\\[0.3em]
a_2^{\text{in}}\\
\end{pmatrix} ,
\label{Eq:ScatteringMatrixExample}
\end{eqnarray}
where the input (output) optical modes are represented by the annihilation operator $a^{\text{in}(\text{out})}$, and the $2\times2$ matrix describes the optical mode mixing process and named scattering matrix, with $r$ and $t$ being the reflectivity and transmissivity satisfying $r^2 + t^2 = 1$ (indicating the optical mixing process corresponds to a unitary evolution).To meet the requirements of our model, we need to create two distinct types of scattering matrices. The first type should explain the mixing process between the modes of the system and the environment, while the second type should describe the mixing process between the different components of the environment. It's important to note that the reflectivity and transmissivity of the two collision cases can be represented by $r_{1(2)}$ and $t_{1(2)}$, respectively. With that in mind, we can now provide a detailed description of the construction process of the scattering matrix for the evolution.

As we introduced before, the evolution of a quantum state through a scattering matrix describes the transition of the state from one moment to another. Allowing the continuous change of a quantum state in the time domain that can be represented as a discrete sequence of states undergoing single or multiple scattering matrices. This indicates that the continuous dynamical map can be equivalently expressed in the following form:
\begin{equation}
\{a\}^{\text{out}}(L)=\mathcal{S}(L)\cdot\{a\}^{\text{in}}(L),
\label{Eq:Mapping}
\end{equation}
where $\mathcal{S}(L) = \prod_{j=1}^{L}\mathcal{S}_{j}$ and the total number of collision routes being $L$. The mode operators describe the entirety of the system. That is, the collection of the annihilation operators, denoted by $\{a\}^{\text{in}(\text{out})}(L)=[a_{An}^{\text{in}(\text{out})},a_{S}^{\text{in}(\text{out})},a_{E,1}^{\text{in}(\text{out})},\cdots, a_{E,L+1}^{\text{in}(\text{out})}]^{\text{T}}$, stores the complete information. Besides, the commutation $[a_k,a^{\dagger}_l]=\delta_{kl}$ holds true for these operators, where the superscript $\text{T}$ represents the transpose, and those subscripts $An$, $S$, and $(E,j)$ denote the ancilla, the system, and the $j$-th environmental mode, respectively.

The $(L+3)$-dimensional matrix $\mathcal{S}_{j}$ is the $j$-th scattering matrix describing the process of mixing all optical modes (both the system-environment and environment-environment are included). Thus, the specific form of the scattering matrix $\mathcal{S}_{j}$ is governed as follows,
\begin{equation}
\mathcal{S}_{j}=
\begin{pmatrix}
1 &0 & 0 & 0 & 0 & 0 \\
0 &r_1e^{i\varphi} & 0 & t_1e^{i\varphi} & 0 & 0 \\
0 &0 & \mathbb{I}_{j-1} & 0 & 0 & 0 \\
0 &-r_2t_1 & 0 & r_1r_2 & t_2 & 0 \\
0 &t_1t_2 & 0 & -r_1t_2 & r_2 & 0 \\
0 &0 & 0 & 0 & 0 & \mathbb{I}_{L-j}
\end{pmatrix},
\label{Eq:SM}
\end{equation}
where $\mathbb{I}_{m}$ is the $m \times m$ identity matrix, and $\varphi$ is the phase shift, which can be realized via a  phase shifter in the optical path.

Our proposal is discussed in the Gaussian regime, and the $N$ bosonic modes can be associated with a tensor product Hilbert space: $\mathcal{H}_{\text{tot}}=\otimes_{k=1}^{N}\mathcal{H}_{k}$. However, the Gaussian system can be described in the quadratic form by defining the corresponding quadrature operators for each optical mode, i.e., $q_{k}=a_{k} + a^{\dagger}_{k}$ and $p_{k}=-i(a_{k} - a^{\dagger}_{k})$.

The quadrature operators obey canonical commutation relations, which read
\begin{equation}
[q_{k},p_{l}]=i\Omega_{kl}, \, \Omega=\bigoplus^{N}_{k=1}\omega, \text{with}\, \,\omega=\begin{pmatrix} 0 & 1\\-1 & 0\end{pmatrix},
\end{equation}
where $\Omega$ is a real, canonical symplectic matrix. Then, a Gaussian state can be fully characterized by the real symmetry matrix, namely covariance matrix (CM) $\sigma$, which satisfies the following Robertson-Schr\"odinger uncertainty relation~\cite{SouzaPRA2015},
\begin{equation}
\sigma+i\Omega\geq0.
\end{equation}
For the quantum state with $N$ bosons, the all matrix elements of CM are specifically defined by second statistical moments of the quadrature operator: $\sigma_{kl} = \langle \lbrace R_{k}, R_{l} \rbrace \rangle/2 - \langle R_k \rangle \langle R_l \rangle$, with a vector of operators $R = (q_1,p_1,\dots,q_N,p_N)^{\text{T}}$ and $\langle\cdot\rangle$ denotes the expectation value.

The CM is a key to our exploration of the system dynamics, and we can choose a convenient way to manage the elements, namely characteristic function formalism $\chi(\lambda)$~\cite{WallsQO1994}. The characteristic function can be obtained through its corresponding density matrix $\rho$, $\chi(\lambda) = \text{Tr}[\rho D(\lambda)]$, where $D(\lambda)$ is the Wely displacement operator defined as $D(\lambda)=\text{exp}\left(\lambda a^{\dagger} - \lambda^{*}a\right)$.  The derivative of the characteristic function in the origin generates symmetrically ordered moments of mode operators. In the formula
\begin{equation}
(-1)^{q}\frac{\partial^{p+q}}{\partial \lambda^{p}_{k} \partial \lambda^{*q}_{l}} \chi(\lambda) \arrowvert_{\lambda = 0}  = \text{Tr}\left\{\rho \left[\left(a^{\dagger}_{k}\right)^{p}a_{l}^{q}\right]_{\text{symm}}\right\}.
\label{Eq:CharaFunMonments}
\end{equation}
Though Eq.~(\ref{Eq:CharaFunMonments}), we can get the corresponding covariance matrix of the input mode of our interest.

Let us discuss our model settings specifically. In our model, each environmental optical mode is set as the identical single-mode squeezed thermal state, and the corresponding input characteristic function has the following form
\begin{equation}
\chi ^{\text{in}}_{E_j}(\lambda) = \text{exp} \left\{ \left(n+\frac{1}{2}\right)\left[\frac{1}{2}\sinh \zeta (e^{-i\phi}\lambda^2 + e^{i\phi}\lambda^{*2}) - \cosh \zeta |\lambda|^2\right] \right\},
\label{Eq:ChaFunE}
\end{equation}
where $n$ is the average thermal photon number, $\zeta$ is the squeezing strength, and $\phi$ denotes the squeezing angle. The parameters $n$, $\zeta$, and $\phi$ are real numbers.

Regarding the input modes of the system and ancilla, they are entangled with each other. Therefore, we consider a two-mode squeezed vacuum state denoted by $\rho_{J}$ to initialize the joint input modes. The TMSV state is a correlated state for two Bosonic modes ($a$ and $b$) and is generated by acting the unitary Bogoliubov transformation $S(\xi) = \text{exp} \{ \xi e^{-i \theta} ab - \xi e^{i\theta} b^{\dagger}a^{\dagger} \}$ on the two-mode vacuum state $| 0_{a},0_{b}\rangle$. Here $\xi$ denotes the squeezing strength and $\theta$ is the squeezing angle. For the sake of simplicity we set $\theta=0$ hereinafter. The TMSV state is a pure state and the corresponding density matrix is $\rho_{An+S} = S(\xi)| 0_{a},0_{b}\rangle \langle 0_{a},0_{b} | S^{\dagger}(\xi)$. As a result, we can obtain the corresponding characteristic function of joint input modes
\begin{equation}
\begin{aligned}
&\chi_{J}^{\text{in}}(\lambda_{An},\lambda_{S}) \\
&= \text{exp}\left(-\frac{|\lambda_{An}|^2 + |\lambda_S|^2}{2} \cosh \xi + \frac{\lambda_{An} \lambda_S + \lambda_{An}^{*} \lambda_{S}^{*}}{2} \sinh \xi\right).
\end{aligned}
\label{inputCharFunc}
\end{equation}
The initial joint input modes of interest are set to be uncorrelated with the environment. As a result, the total input characteristic function of this collision model $\chi_{T}^{\text{in}}(\vec{\lambda})$ is calculated by a product form
\begin{equation}
\chi_{T}^{\text{in}}(\vec{\lambda}) = \chi_{J}^{\text{in}} \times \prod_{j=1}^{L} \chi_{E_j}^{\text{in}},
\label{ChaFunIn}
\end{equation}
where $\vec{\lambda} = [\lambda_{An},\lambda_{S},\lambda_{E_1},\dots,\lambda_{E_j},\dots,\lambda_{E_{L+1}}]$, with $\lambda_{E,j}$ is the characteristic function of the $j$-th environmental mode. In addition, as we mentioned before, the scattering matrix $\mathcal{S}(L)$ can connect the characteristic function of the input modes and output modes after they have undergone $L$ times collision routes, shown in the following~\cite{LiPRA2020,JinNJP2018}
\begin{equation}
\chi^{\text{out},L}_{T}(\vec{\lambda}) = \chi^{\text{in}}_T[\mathcal{S}^{-1}(L)\vec{\lambda}].
\label{ChaFunINOUT}
\end{equation}
To achieve the desired output characteristic function, one approach is to partially trace over all other modes by substituting a specific vector~\cite{WangPR2007}, e.g., substituting $\vec{\lambda} = [\lambda_{An},\lambda_{S},0,\dots,0]$ into above equation to obtain the characteristic function of the joint mode. Thus, we will have the following form
\begin{equation}
\begin{aligned}
&\chi_{J}^{\text{out},L}(\lambda_{An},\lambda_{S})\\
=&\text{exp}\{\left(n+\frac{1}{2}\right)\left[\frac{1}{2}\sinh\zeta\left(e^{-i\phi}\sum_{l=3}^{L}(C_{l,2}\lambda_{S})^2+e^{i\phi}\sum_{l=3}^{L}(C_{l,2}^{*}\lambda_{S}^{*})^2 \right) \right.\\
&\left.+\cosh\zeta\sum_{l=3}^{L}|C_{l,2}|^2|\lambda_{S}|^2\right] \}\cdot\chi^{\text{in}}_{J}(\lambda_{An},C_{2,2}\cdot\lambda_{S}),
\end{aligned}
\label{OutputChaFunjuti}
\end{equation}
where $C_{l,k}$ indicates the matrix element in the $l$-th row and $k$-th column of the  inverse of scattering matrix $\mathcal{S}^{-1}(L)$. With the help of Eq.~(\ref{Eq:CharaFunMonments}), we are able to obtain the expectation values of the operators we are interested in.

\section{measurement of non-Markovianity}
\label{Sec:NMmeasure}
As we proceed with this section, we briefly review the measure of non-Marovianity based on Gaussian steering. In addition, we will also introduce the measure based on the violation of the Gaussian channel's divisibility for benchmark comparison purposes.

\subsection{Gaussian steering to witness and measure non-Markovianity}
EPR steering is a form of non-local correlation that allows one part of a bipartite quantum system to steer the other part through a local measurement. In the context of this scenario, we have Alice and Bob as two experimenters who share a bipartite quantum state, represented by $\rho_{AB}$. Alice then proceeds to perform a set of local measurements, $\mathcal{M}$, on her side of the state. With the measurements $R_A$ and $R_B$ ($R_A \in \mathcal{M}$, and $R_B$ is arbitrary) and their outcomes $r_A$ and $r_B$, when the joint probability $P(r_A,r_B|R_A,R_B,\rho_{AB}) = \sum_{\nu} \mathscr{P}_{\nu} \mathscr{P}(r_A|R_A,\lambda)P(r_B|R_B,\rho_{\nu})$ is violated by at least one measurement pair $R_A$ and $R_B$, Alice can effectively steer the types of the states that Bob can have access to in his side. Here $\mathscr{P}_{\nu}$ across all measurements, $\mathscr{P}_{\nu}$ and $\mathscr{P}(r_A|R_A,\nu)$ are arbitrary probability distributions and $P(r_B|R_B,\rho_{\nu})$ is the hidden variable that Bob holds \cite{Wiseman2007PRL, Kogias2015PRL}.

We narrow down our discussion to the Gaussian realm. The joint mode of the system and ancilla of our study is a Gaussian state, which the joint state can be described by the covariance matrix $\sigma_{J}$, and the measurement $\mathcal{M}$ is also Gaussian. With the help of this work, see Ref.~\cite{Kogias2015PRL}, the measure of the Gaussian steering can be defined.

Our studies aim to address two distinct scenarios. The first scenario involves the system mode having the ability to steer the ancilla, and the corresponding measure is \textit{Gaussian $S \rightarrow An$ steerability}
\begin{equation}
\mathcal{G}^{S\to An} = \text{max} \{ 0,\frac{1}{2} \ln \frac{\text{det}[V_{S}]}{\text{det} [\sigma_{J}]} \}.
\label{Eq:GaussianSteeringS}
\end{equation}
On the other hand, the second case pertains to the opposite scenario where the system mode can be steered by the ancilla, which is referred to as \textit{Gaussian $An \rightarrow S$ steerability}
\begin{equation}
\mathcal{G}^{An\to S} = \text{max} \lbrace 0,\frac{1}{2} \ln \frac{\text{det}[V_{An}]}{\text{det} [\sigma_{J}]} \rbrace.
\label{Eq:GaussianSteeringAn}
\end{equation}
where $\sigma_{J}$ having the form
\begin{equation}
\sigma_{J} =
\begin{pmatrix}
V_{An} & V_{J} \\
V_{J}^{\text{T}} & V_{S}
\end{pmatrix},
\label{CovarianceMatrix}
\end{equation}
where $V_{An}$ and $V_S$ are the covariance matrices corresponding to the ancilla and the system, and $V_{J}$ is the correlation matrix of this joint mode. The positive value of the measure implies the steerability, e.g., when $\mathcal{G}^{S \to An}$ is positive, it means that the system mode can steer the ancilla mode. The detailed expressions of those matrices can be referred to in Appendix~\ref{Sec:Appendix A}.

In the recent work~\cite{FrigerioPRA2021}, Frigerio and his collaborators exploit Gaussian steering to probe non-Markovianity through the quantum Brownian motion channel. It shows that if Gaussian steering is not strictly monotonically decreasing during the evolution, we can say that the channel is non-Markovian, which means that the time derivative of the steering measure is positive. This is due to the fact that Gaussian steering, as a kind of quantum resource, cannot exceed its initial value under generic local Gaussian operations in a general CPTP map. This means that if the Gaussian channel satisfies CP divisibility, the steering will not increase; otherwise the Gaussian steering will show some non-monotonical behaviors. In this work, we restrict our discussions to the field of Gaussian channels and operations. As a result, in the collisional picture of the Gaussian system, the non-Markovian dynamics can also be defined based on the expression given in Ref.~\cite{FrigerioPRA2021},
\begin{equation}
\mathcal{D}_{j} = \text{max} \lbrace0,\mathcal{G}^{\to}[\sigma_{J}(j)]- \mathcal{G}^{\to}[\sigma_{J}(j-1)]\rbrace>0,
\label{eq:derivative}
\end{equation}
where $\sigma_{J}(j)$ is the joint covariance matrix after $j$ times collision route, and $\to$ is a general index that can indicate different steering scenarios.

Hence, the measurement of non-Markovianity for a Gaussian channel can be defined as follows:
\begin{equation}
\mathcal{N}_{\text{GS}}(L) = \sum_{j=1}^{L} \mathcal{D}_j.
\label{eq:nm}
\end{equation}
The above definition can be interpreted as follows: the degree of non-Markovianity can be qualified by summing up the violation of Markovianity that occurs during the collision process \cite{Chen2017PRL,FrigerioPRA2021}.

\subsection{Measuring non-Markovianity of Gaussian channel based on violation of the CPTP map divisibility}
\label{SEC:CPTP}

We now introduce an extensively used measurement of non-Markovianity for Gaussian channels, proposed by Torre {\it et al.} in Ref.~\cite{TorrePRL2015}, which is based on quantifying the degree of violation of dynamical divisibility.

Under the collision model's framework, the time-evolution process of the system can always be described by the dynamical map $\mathcal{E}_{j}[\cdot]$, i.e., after the $j$-th time collision route occurring, the output covariance matrix of the system can be
\begin{equation}
   \sigma_{S}(j)=\mathcal{E}_{j}[\sigma^{\text{in}}_{S}].
\end{equation}
Then, the stroboscopic property of the collision model ensures that we can split its dynamical map as follows,
\begin{equation}
    \mathcal{E}_{j}=\Phi_{j,j-1}\circ\mathcal{E}_{j-1},
\end{equation}
where $\Phi_{j,j-1}$ corresponds an intermediate process and the `$\circ$' denotes the composition of the map. The Markovian dynamic suggests that the dynamical map is always divisible, indicating that $\Phi_{j,j-1}$ is CPTP for all collision processes. However, nondivisibility can occur in the presence of certain intermediate processes, where $\Phi_{j,j-1}$ fails to obey CPTP. This implies the non-Markovian dynamic - the other side of the coin.

As for a generic Gaussian channel, the dynamical map $\mathcal{E}_{j}[\cdot]$ has the following form: $\sigma_{S}(j)=\mathcal{E}_{j}[\sigma^{\text{in}}_{S}]=X_j \sigma_{S}^{\text{in}} X_{j}^{T} + Y_{j}$~\cite{TorrePRL2015}, with $X_{j}$ and $Y_{j}$ are $2 \times 2$ matrices. Then, we can find that the sufficient and necessary conditions for the intermediate process $\Phi_{j,j-1}$ to satisfy the CPTP condition is the semi-positive definiteness of the following $2 \times 2$ matrix,
\begin{equation}
\mathcal{F}_{j} = Y_{j,j-1} - \frac{i}{2} \Omega + \frac{i}{2} X_{j,j-1} \Omega X_{j,j-1}^{\text{T}},
\label{Eq_juzhen}
\end{equation}
where $X_{j,j-1} = X_{j}X_{j-1}^{-1}$, $Y_{j,j-1} = Y_{j} - X_{j,j-1}Y_{j-1}X_{j,j-1}^{\text{T}}$, and $\Omega$ has been given before. The detailed expressions of the above matrices and relevant inferences can be found in Appendix~\ref{Sec:Appendix B}.  The eigenvalues of $\mathcal{F}_{j}$ are related to the CPTP property of $\Phi_{j,j-1}$: the negative eigenvalue of $\mathcal{F}_{j}$ leads to a non-CPTP for $\Phi_{j,j-1}$ and thus for the non-Markovianity of Gaussian channel with vacuum environment can be quantified by~\cite{JinNJP2018}
\begin{equation}
\mathcal{N}_{\text{CPTP}}(L) = \sum_{j=2}^{L} \sum_{m = \pm} \frac{|\nu_{j,m}| - \nu_{j,m}}{2},
\label{Eq:NCPTP}
\end{equation}
where $\nu_{j,m}$ are the eigenvalues of $\mathcal{F}_{j}$. In addition, from the expression of $\nu_{j,m}$ shown in Appendix~\ref{Sec:Appendix B} we can find that quantifying non-Markovianity using this measurement is related to two conditions: the structure of the scattering matrix and the states of the environment. Both steering cases have the same scattering matrix and environment states in our collision model. Thus, we can consider the arbitrary side of steering as the initial input system mode to quantify the non-Markovianity of the dissipation channel.

\section{Results and Discussions}
\label{Sec:Result}
\subsection{Measuring non-Markovianity by Gaussian steering}
\label{GaussianSteering}

\begin{figure}[!htpb]
  \includegraphics[width=\linewidth]{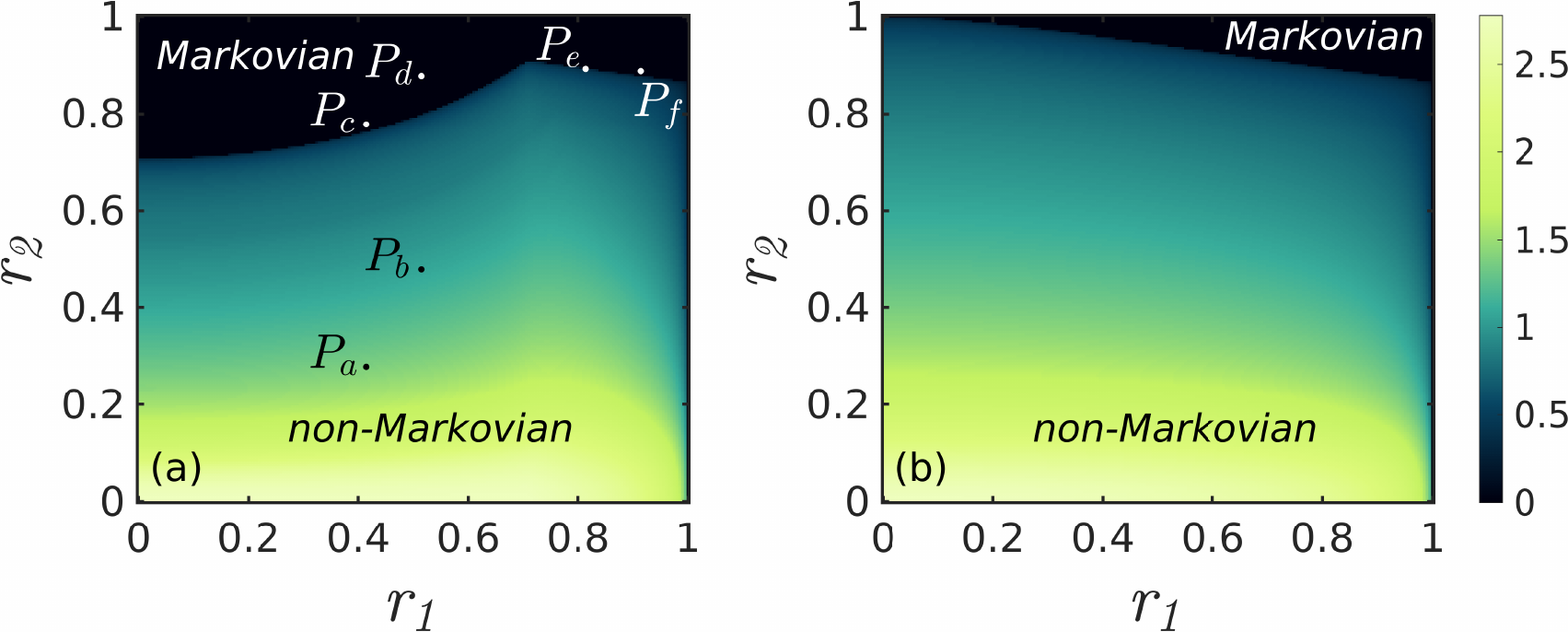}
  \caption{\label{FIG:NMQS} The Markovian and non-Markovian regions which are quantified by the Gaussian steering $\mathcal{N}_{\text{GS}}(L)$ in the $r_1$-$r_2$ plane. The environmental modes are vacuum states.  In both panels  the dissipation channel acts on the system mode, but for (a) the system can steer the ancilla, and for (b), the system will be steered by the ancilla. The phase shift is always $\varphi=0$. The number of collisions in each case is $L=250$.}
\end{figure}

We start our study with the results of non-Markovianity degree in the reflectivities parameter space. As shown in Fig.~\ref{FIG:NMQS}, the non-Markovianity degree of dissipative evolution process in the $r_1$-$r_2$ plane is quantified by Gaussian steering for the two cases with the environmental states are all initialized vacuum states, as previously mentioned~\cite{footnote1}.

\begin{figure*}[!hptb]
  \includegraphics[width=1\linewidth]{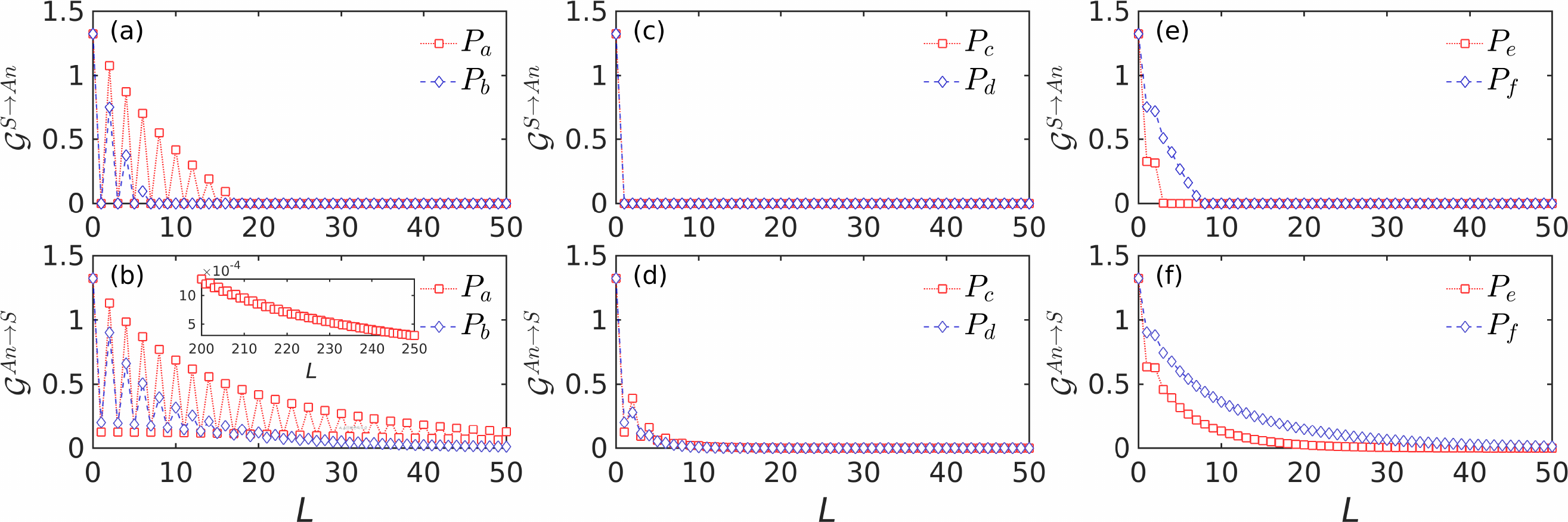}
  \caption{\label{FIG:QSevolution} The Gaussian steerabilities $\mathcal{G}^{\to }$ in Eqs.~(\ref{Eq:GaussianSteeringS}) and (\ref{Eq:GaussianSteeringAn}), as functions of the number of steps $L$ in the different scenarios: the first and second rows correspond to the cases that the system can steer the ancilla, $\mathcal{G}^{S \to An}$, and the system can be steered by the ancilla, $\mathcal{G}^{An \to S}$, respectively. We have chosen six sets of parameters, denoted as $P_{\beta} = (r_{1},r_{2})$ with $\beta=a,b,c,d,e$ and $f$ for each case. The specific parameter settings are $P_{a}=(0.4,0.3)$, $P_{b}=(0.5,0.5)$, $P_{c}=(0.4,0.8)$, $P_{d}=(0.5,0.9)$, $P_{e}=(0.8,0.9)$ and $P_{f}=(0.9,0.9)$, and these parameter points are also have been labeled in Fig.~\ref{FIG:NMQS}(a).}
\end{figure*}

Our primary attention is directed towards the extreme case in the parameter space. It has been observed that if the reflectivity is always fixed at $r_{2}=1$, then the dynamics are Markovian, irrespective of the value of $r_1$ for both two cases. The reason for this is quite straightforward. When $r_2=1$, there is no mixing of information between environmental modes, and quantum correlation cannot be established. This is illustrated in Fig.~\ref{FIG:Model}, in which we can see that the system mode and the $E_{j+1}$ mode in the $j$-th collision route are unable to build any quantum correlation. As a consequent, in each collision route, the system mode will always interact with a new environmental mode that does not carry any of its historical information. As a result, information flows only in one direction - from the system mode to the environment - with no backflow of information. This behavior is a hallmark of a standard Markovian process. Furthermore, It is also worth noting the other extreme case, where the dynamics of the system also exhibit Markovian behavior as the reflectivity approaches $r_{1}\approx 1$. However, this type of dynamical process actually corresponds to unitary evolution, i.e., the system does not actually interact with environmental modes, which is a closed quantum system.

Within the non-extreme regions, when we fix the $r_{1}$ and vary $r_{2}$, the degree of non-Markovianity of the system decreases as $r_{2}$ increases. It can be attributed to the fact that when the exchange of information between the system and the environment is constant, a decrease in $r_{2}$ leads to a higher flow of information back into the system, hinting at the presence of stronger non-Markovian dynamics.

In addition, upon comparing the results presented in Fig.~\ref{FIG:NMQS}, it is evident that the Markovian regions in the parameter space differ for those two scenarios. Specifically, the Markovian region of Fig.~\ref{FIG:NMQS}(a) is larger than that of Fig.~\ref{FIG:NMQS}(b). In the region of $r_1 \lessapprox 0.7$ and $0.7 \lessapprox r_2 \leq1$, it can be observed that the case in which the system can steer the ancilla exhibits the non-Markovian region while the other case still maintains its Markovian region.

To analyze the specific differences between the two scenarios, we opted to choose certain parameter points, denoted as $P_{a},\cdots, P_{f}$ in Fig.~\ref{FIG:NMQS}, and showcase their corresponding Gaussian steering $\mathcal{G}^{\to}$ dynamic behavior in Fig.~\ref{FIG:QSevolution}.

We start our discussions with the parameter points $P_{a}$ and $P_{b}$. It can be found that both cases can exhibit non-Markovian properties in Fig.~\ref{FIG:NMQS}(a)-(b), and this observation is supported by the results shown in Fig.~\ref{FIG:QSevolution}(a)-(b). The time-evolution quantum steering exhibits non-monotonic behavior and hints the occurrence of information backflow. However, Upon closer examination, it becomes apparent that even though the parameters of the dissipation channel are identical in the two cases, the steering behavior in Fig.~\ref{FIG:QSevolution}(b) shows a significant difference. More specifically, the quantum steering does not come close to disappearing until a sufficient number of collisions have taken place, especially at $P_{a}$, i.e., $L=250$. Moreover, in Fig.~\ref{FIG:QSevolution}(a), we can observe the Gaussian steering experiences the sudden change during its evolution process, i.e., the value of Gaussian steering will suddenly drop to zero and promptly recover at the next collision. This pattern continues until reaching a specific step until the steering does not recover any longer. Then, we proceed with our discussions on the situation of the parameter points $P_e$ and $P_f$ in Fig.~\ref{FIG:NMQS}, which correspond to Fig.~\ref{FIG:QSevolution}(e) and Fig.~\ref{FIG:QSevolution}(f), respectively, and all of which correspond to Markovian dynamic in both dissipation cases. In either case, the Gaussian steering evolution is monotonically decreasing, which is a characteristic property of Markovian processes. However, there is a difference in the evolution process between $\mathcal{G}^{An \to S}$ and $\mathcal{G}^{S \to An}$. The time-evolution of $\mathcal{G}^{An \to S}$ approaches zero asymptotically, while in the case of $\mathcal{G}^{S \to An}$, it suddenly becomes zero at a certain collision step without any possibility of recovery.

\begin{figure}[!htpb]
  \includegraphics[width=0.94\linewidth]{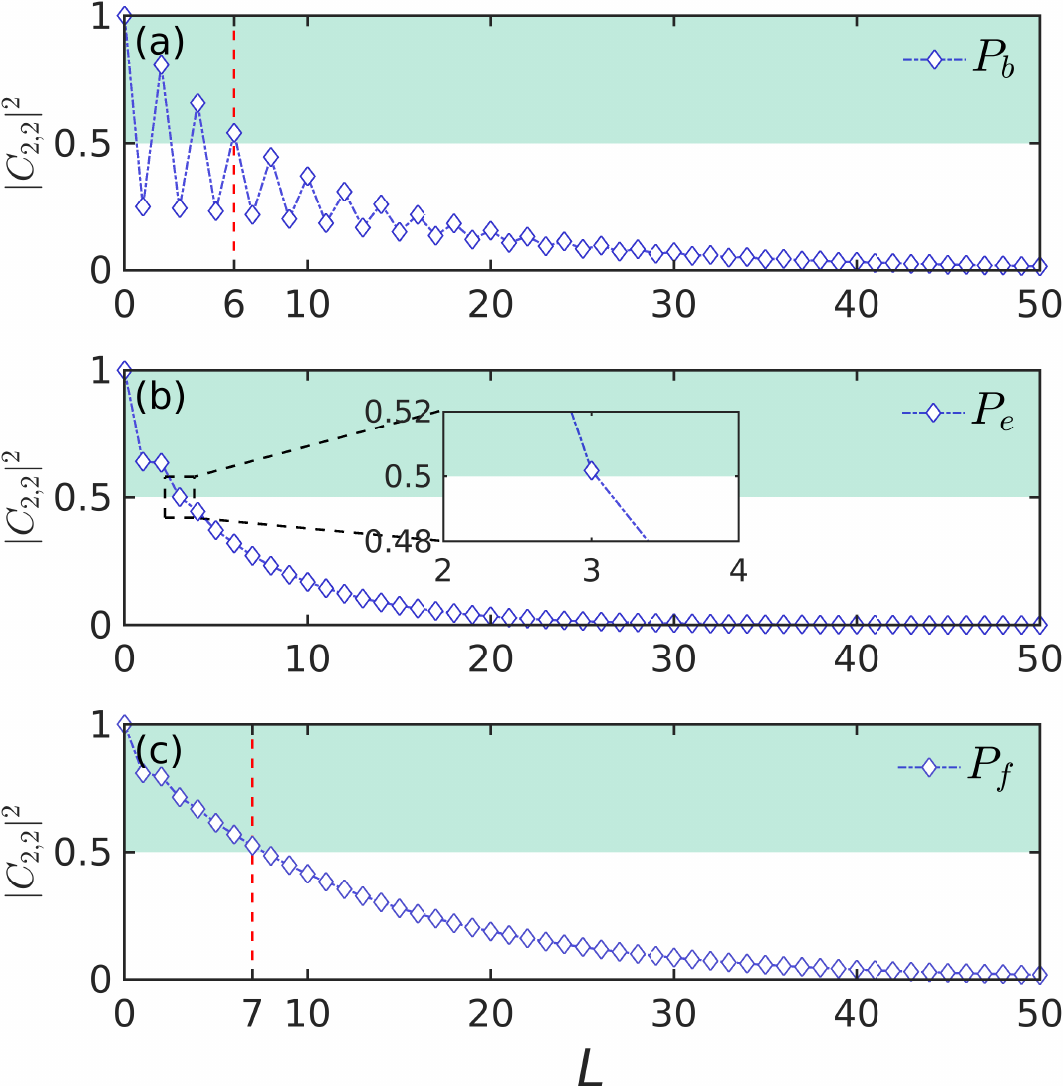}
  \caption{\label{FIG:C11} Variation of the matrix element $|C_{2,2}|^2$ with the number of collisions $L$. Each panel (a), (b), and (c) corresponded to the parameter points $P_b$, $P_e$, and $P_f$ of the $\mathcal{G}^{S \to An}$ case in Fig.~\ref{FIG:QSevolution}, respectively.}
\end{figure}

In those previously discussed parameter points, both cases arrive at similar conclusions regarding the properties of the system dynamics. However, the discrepancy between the two lies in the specific details of the steering evolution over time, leading to varying degrees of non-Markovian within the system dynamics. Then, we will discuss the middle two panels of Fig.~\ref{FIG:QSevolution}, in which the system dynamics show totally different properties. It can be found that in Fig.~\ref{FIG:QSevolution}(c), the Gaussian steering can suddenly become zero when $L=2$ for $\mathcal{G}^{S \to An}$, hints the Markovian properties, while for the $\mathcal{G}^{An \to S}$ in Fig.~\ref{FIG:QSevolution}(d), the dynamic is non-Markovian with a non-monotonic decreasing behavior of quantum steering.

We recall our model settings here, including the dissipation channel and its corresponding parameters; all settings are the same in both scenarios; the only difference is whether the system mode is the steering or steered parts. Therefore, it is interesting to note that the dynamic behavior of Gaussian steering is significantly different in both cases.

To gain better insights into the discrepancies we mentioned above, it is worth recalling the definition of the measure of non-Markovianity: the dynamics of Gaussian steering exhibit a non-monotonically decreasing behavior. However, it is crucial to note that this definition is always based on the fundamental assumption that the model is always steerable. To this end, discussing the conditions for when the system or the ancilla is steerable will be meaningful.

We start our discussions with the system as the steering part case.  In accordance with Eq.~(\ref{Eq:GaussianSteeringS}), steerability can be achieved when satisfying $\text{det} [V_S] > \text{det} [\sigma_{J}]$. By refining this condition, we can derive a more dynamic expression that only pertains to the matrix element for the inverse of the scattering matrix $\mathcal{S}^{-1}$ after the $j$-th collision,
\begin{equation}
|C_{2,2}(j)|^2 > 1 - \frac{1}{2(n+1)},
\label{Eq:SysSteerable}
\end{equation}
where $n$ is the average thermal photon number of the environment states we have stated before. The above condition will be reduced into $|C_{2,2}(j)|^2 > 1/2$ for when the environment states are initialized as the vacuum mode ($n=0$).

When considering the scenario where the system is the steered part, it is worth noting that the expression of the steerability condition becomes intricate. The steerable conditions are related to the initial setting of the system state's squeezing strength $\xi$, the environment states' squeezing strength $\zeta$, and also the average thermal photon number. In the case of the environment modes are thermal state, i.e., $\zeta=0$, we can obtain the general form of the condition,
\begin{equation}
|C_{2,2}(j)|^2 > \frac{2n\cosh\xi}{(2n+1)\cosh\xi-1}.
\label{Eq:AnSteerable}
\end{equation}
To provide a more comprehensive understanding, we have included the detailed derivations of the above inequalities in Appendix~\ref{Sec:Appendix C}. Additionally, it is important to note that the above condition shown in Eq.~(\ref{Eq:AnSteerable}) will always hold when $n=0$, and it implies that the ancilla mode has the capability to steer the system mode at all times.

In order to verify our previous conclusions, we present the scenario where the system mode will be the steering part. Specifically, we analyze the variation of $|C_{2,2}(j)|^2$ with respect to the collision times. The results are given in Fig.~\ref{FIG:C11}.

We highlight three distinct parameter settings in our following discussions, illustrated in Figs. \ref{FIG:C11}(a)-(c) that correspond to the points $P_{b}$, $P_{e}$, and $P_{f}$ as shown in Figs. \ref{FIG:QSevolution}(a) and \ref{FIG:QSevolution}(e). The three panels are all divided into two regions; the green part is to be the regime of $|C_{2,2}(j)|^2>1/2$, and the residual part is to be white.

We start our discussions with the $P_{b}$ case. It can be clearly found that when the number of collisions is restricted to $j\leq6$, the value of $|C_{2,2}(j)|^2$ varies iteratively with the collision process around $|C_{2,2}(j)|^2=1/2$. This phenomenon implies that $\mathcal{G}^{S\to An}$ will change suddenly as the collision process occurs, and this coincides with what we observed in Fig.~\ref{FIG:QSevolution}(a). Similarly, we can compare the dynamic behaviors of $P_{e}$ and $P_{f}$ shown in Fig.~\ref{FIG:C11} and Fig.~\ref{FIG:QSevolution}. We find the numerics also suggest that in the case of $\mathcal{G}^{S\to An}$, the dissipative process will have a significant impact on the survival of Gaussian steering: due to the influence of dissipation, the system mode may lose its ability to steer the ancilla, which can be observed from the Markovian dynamic results. On the other hand, the asymmetric property of quantum steering can lead to the different non-Markovian regions for the two cases, as can be observed in Fig.~\ref{FIG:NMQS}. It also means that inaccurate results will be obtained if the system functions as the steering part while the dissipative channel is in action, as this can affect the measure of non-Markovian.

\subsection{Sudden death and sudden birth of Gaussian steering}

\begin{figure}[!htpb]
  \includegraphics[width=0.95\linewidth]{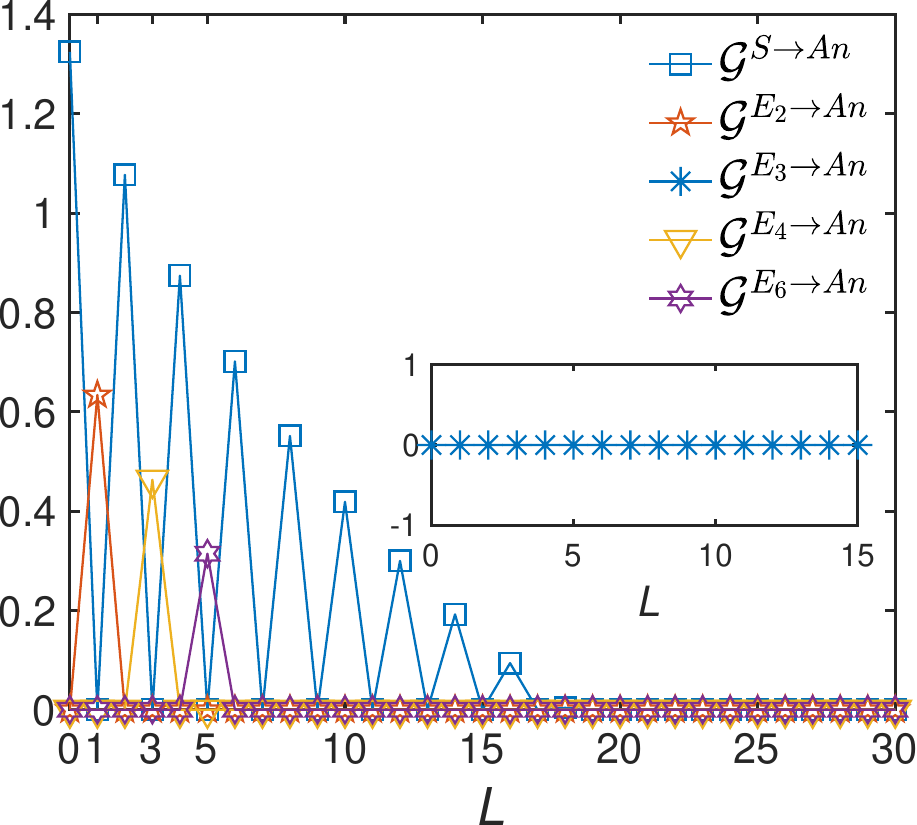}
  \caption{\label{FIG:QStrans} Gaussian steerability $\mathcal{G}^{\to}$ as a function of collision time for $r_{1}=0.4$, $r_{2}=0.3$. }
\end{figure}

As we have seen in the previous section, Gaussian steering $\mathcal{G}^{S\to An}$ can exhibit sudden changes in certain parameter cases, as shown in Fig.~\ref{FIG:QSevolution}(a). To gain a deeper understanding of this behavior, we also examined it from a mathematical perspective, as depicted in Fig.~\ref{FIG:C11}(a). However, further investigation is required to uncover the underlying physics fully. To this end, we consider to focus on the details of the evolution of the entire model.

Our all-optical collision model is designed to take account of all the necessary information about the auxiliary, system, and environment parts throughout the numerical simulation. Although we mainly concentrate on the system component in the past discussion, the model encompasses all aspects to comprehensively represent the system's evolution. This implies that quantum correlation between the subsystems will be retained in the overall system, even though it may be transported, delocalized, or undergo some other changes. This also makes it possible to investigate the sudden changes in Gaussian steering more specifically.

The observed sudden changes in Gaussian steering can be attributed to the transfer of quantum steering from the system to other subsystems. In light of this, we show the Gaussian steerability to the ancillary mode as a function of the number of collisions in different subsystems. The numerics are shown in Fig.~\ref{FIG:QStrans}. It can be seen that in the non-Markovian region ($r_{1}=0.4$, $r_{2}=0.3$), the time evolution of the Gaussian steering $\mathcal{G}^{S\to An}$ shows the sudden changes. After completing the first round of the collision route $(L=1)$, the Gaussian steerability $\mathcal{G}^{S\to An}$ disappeared. However, at the same time, the second environment mode $E_{2}$ became capable of steering the ancilla, with $\mathcal{G}^{E_{2}\to An}>0$. This can be understood as in the first round of collision, the steerability transports into the first environment mode leading to the vanishing of $\mathcal{G}^{S\to An}$ and a nonzero $\mathcal{G}^{E_1\to An}$ . Then the collision takes places between the environment modes $E_{1}$ and $E_{2}$ at the BS (with the reflectivity $r_2$). The small reflectivity $r_2$ allows for an almost exchange between the two environment modes. As a result, the steerability of environmental modes is passed on from $E_{1}$ to $E_{2}$.

To corroborate our analysis, we also include data regarding the variation of the Gaussian steerability of the environment modes $E_{4}$ and $E_{6}$ to the ancilla mode with the number of collisions. Upon analyzing the data, it is evident that the Gaussian steerability is transported from the environment modes to the system mode at $L=2,4,\cdots$, and is stored in the environment modes at $L=1,3,5,\cdots$. In addition, Upon analyzing our collision process, we can discover that during the second round of collision route occurrence, the system mode interacted with the environment mode $E_{2}$ first. At this point, the steerability had already been transferred to the system mode. As a result, the subsequent interaction between environment modes $E_{2}$ and $E_{3}$ did not transfer the steerability to $E_{3}$ mode. This point can be verified by the inset of Fig.~\ref{FIG:QStrans}.

\begin{figure}[!htpb]
  \includegraphics[width=0.95\linewidth]{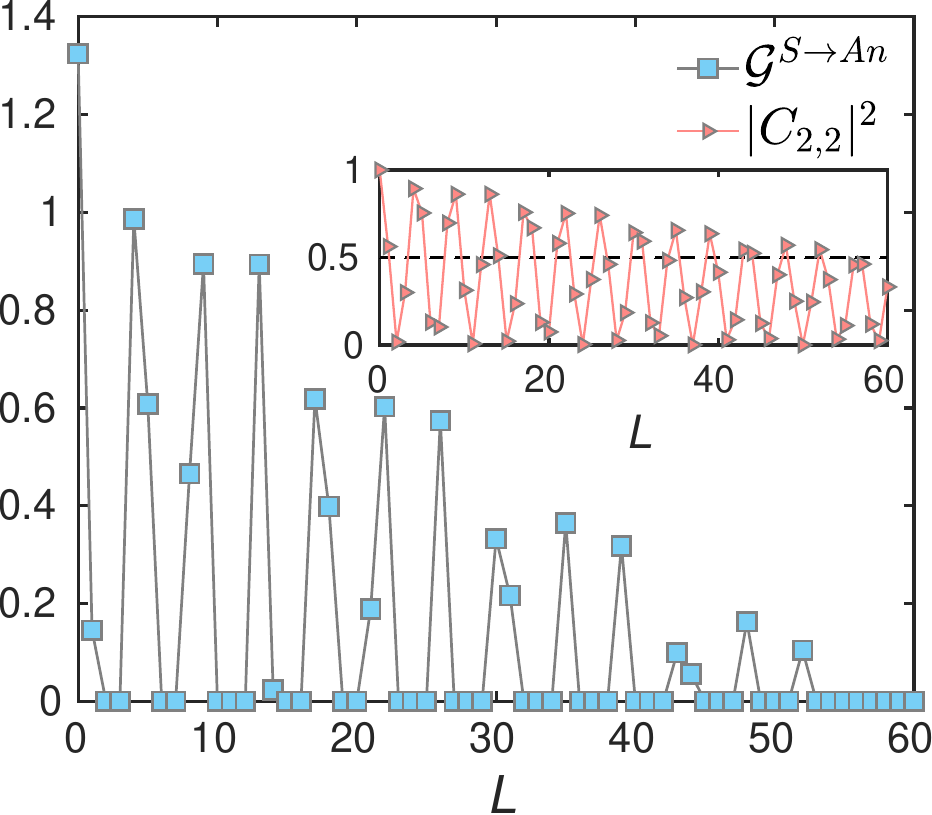}
  \caption{\label{FIG:Sudden} Gaussian steerability $\mathcal{G}^{S\to An}$ and $|C_{2,2}|^2$ (inset) as a function of collision time. The parameters are $r_{1}=0.75$, $r_{2}=0.15$ and $\varphi=\pi$.}
\end{figure}

Indeed, the sudden-change behavior of the Gaussian steering mentioned above can be interpreted as the occurrence of what is commonly referred to as the sudden death or sudden birth of Gaussian steering~\cite{HuEPJC2021,DengNPJQI2021,RosarioPLA2022,LiSR2023}. To get further insights, we introduce a phase shift in the system mode, e.g., $\varphi=\pi$ in the scattering matrix (\ref{Eq:SM}), and investigate the time-evolution of the joint system. In Fig.~\ref{FIG:Sudden}, one can see that the Gaussian steering demonstrates a recurring pattern of transitions, it disappears suddenly at some step and remains vanishing for a finite number of steps then resurfaces suddenly, so on and so forth.

\begin{figure*}[!htpb]
  \includegraphics[width=0.98\linewidth]{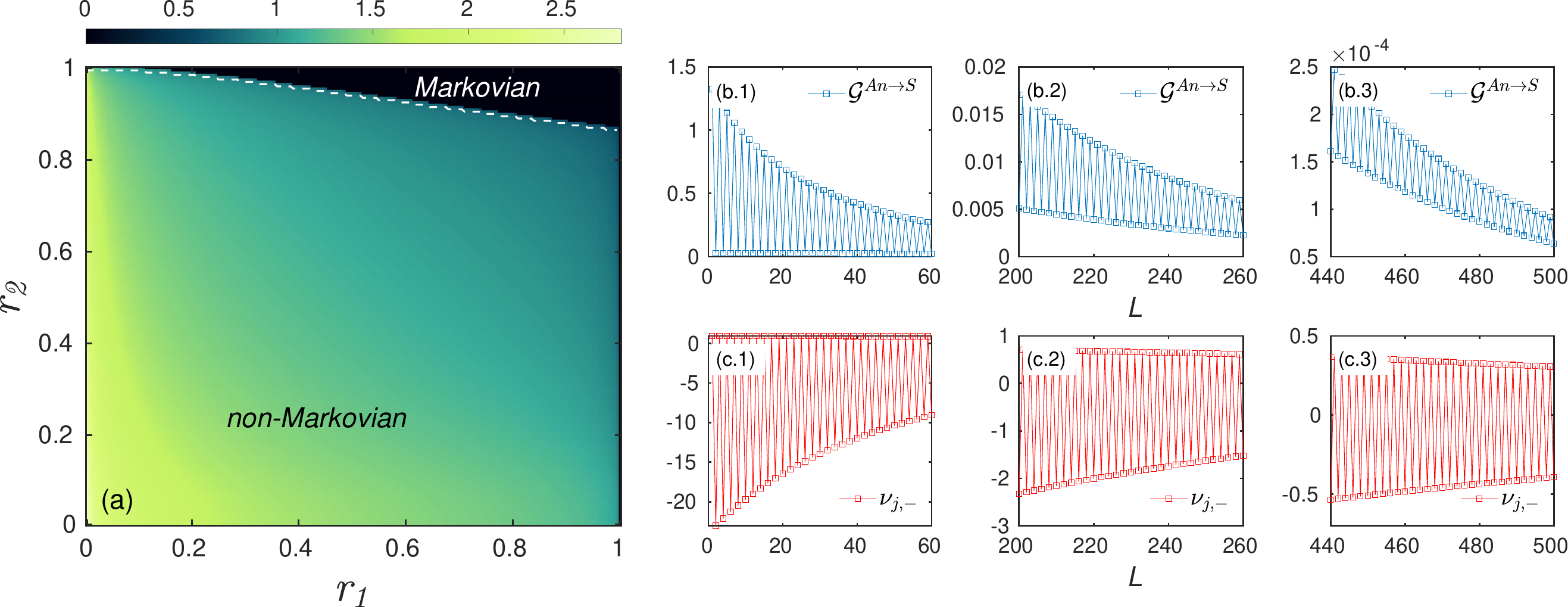}
  \caption{\label{FIG:NMCPTP} (a) The Markovian and non-Markoian regions in the $r_1-r_2$ plane, which is measured by the violation of the CPTP map divisibility $\mathcal{N}_{\text{CPTP}}(L)$ in Eq.~(\ref{Eq:NCPTP}). The white dashed line corresponds to the boundary of the Markovian and non-Markovian regions in Fig.~\ref{FIG:NMQS}(b). The number of collisions is $L=250$. (b.1)-(b.3) The time evolution of Gaussian steering $\mathcal{G}^{An\to S}$ in the different intervals of the number of collision. (c.1)-(c.3) The variation of $\nu_{j,-}$ with the different intervals of the number of collision. The environmental optical modes are initialized to be the vacuum states, and the other parameters are specified as  $r_{1}=r_{2}=0.2$ in (b) and (c). }
\end{figure*}

\subsection{Comparisons of results with the measure based on  the violation of CPTP map divisibility}
We identified the reasons for the distinct dynamic processes exhibited by the two steering scenarios, $\mathcal{G}^{An\to S}$ and $\mathcal{G}^{S\to An}$ in the previous section. However, we have not yet determined which of the scenarios will provide more accurate results for distinguishing the non-Markovian dynamics. Therefore, we introduce the other measures for comparison and to validate the previous results. This section delves into the non-Markovianity property by adopting the concept presented in Ref.~\cite{TorrePRL2015}. As mentioned in Sec.~\ref{SEC:CPTP}, we have already introduced the detailed expression of the CPTP map divisibility.

We show the non-Markovianity based on the violation of the CPTP map divisibility in Fig.~\ref{FIG:NMCPTP}(a). It can be observed that the boundary between Markovian and non-Markovian regions is similar to the $\mathcal{G}^{An\to S}$ steering case illustrated in Fig.~\ref{FIG:NMQS}(b). To further comparison, we have shown the boundary in Fig.~\ref{FIG:NMQS}(b) using the white dashed line in Fig.~\ref{FIG:NMCPTP}(a), and the results can be found to have matched well. Taking our discussion a step further,
we show in details the time-evolution of those two measures in Figs.~\ref{FIG:NMCPTP}(b.1)-(b.3) and \ref{FIG:NMCPTP}(c.1)-(c.3), respectively.

Recall that the condition for the violation of CPTP map divisibility: at least one of the eigenvalues $\nu_{j,\pm}$ of the matrix given in Eq.~(\ref{Eq_juzhen}) is negative. However, we are able to prove that the eigenvalue of $\mathcal{F}_{j}$, $\nu_{j,+}$ is always zero when the environment modes is in vacuum state (c.f. Appendix \ref{Sec:Appendix B} for details). In this way, we only need to focus on the eigenvalue $\nu_{j,-}$, and its dependence on the collision times. In Figs.~\ref{FIG:NMCPTP}(c.1)-\ref{FIG:NMCPTP}(c.3), we provide numerical illustrations of these variations. We set the total number of collision routes to be $L=500$ and focus on three segments of the time evolution: (i) the initial stage from $L=0$ to $60$; (ii) the intermediate stage ranging from $L=200$ to $260$, and (iii) the final stage ranging from $L=440$ to $550$.

In the initial stage, both scenarios exhibit equivalent properties: the negative eigenvalue always exists and the Gaussian steering shows the nonmonotonic evolution. However as the system evolves to the intermediate stage, although the Gaussian steering $\mathcal{G}^{An\to S}$ still shows the nonmonotonic evolution, it becomes very small with the magnitude $10^{-4}$. This is due to Gaussian steering as a resource that may lost during the time-evolution. In principle, in order to characterize the non-Markovianity one should initialize the system and ancilla modes in the maximally entangled state. However, the negative eigenvalue is still visible even if in the late stage of the evolution since the on the violation of CPTP map divisibility is only determined by the intrinsic property of the Gaussian channel.

\section{Summary}
\label{Sec:Summary}

In this paper, we investigate Gaussian steering as a measure to witness non-Markovian dynamic processes. Our work differs from previous studies in that it introduces an all-optical scheme to simulate a quantum collision model, which is considered an effective method for studying the dynamics in open quantum systems. Under the framework of collision models, the continuous time-evolution process can be replaced by the stroboscopic interactions in the spatial dimension. This allows us to implement the different dynamic evolution processes relatively easily and even discuss dynamical transitions.

Therefore, by manipulating the transmission or reflectivity of the beam splitters, we can alter the non-Markovian dynamic processes to Markovian dynamic processes. Our approach involves two steering scenarios where the system mode can steer the ancilla mode, denoted as $\mathcal{G}^{S\to An}$, and where the ancilla mode can steer the system, denoted as $\mathcal{G}^{An\to S}$.
While both scenarios can provide the boundaries between Markovian and non-Markovian dynamics in parameter space, there are significant differences in their results. Specifically, when examining the dynamic processes, we observe that the first scenario can lead to non-trivial changes in the non-Markovian region, such as sudden death and sudden birth.

To address the questions raised, we analyze the conditions for the existence of Gaussian steering; the difference in those conditions is the reason for the different non-Markovian measure results between the two scenarios. We identify the asymmetric properties of Gaussian steering as the underlying reason for inducing distinct recognition of these dynamics. Additionally, we study the changes in quantum non-local correlation in the environment modes to explain sudden changes in Gaussian steering. In order to verify the suitable steering scenario, we compare the results of Gaussian steering with data based on the violation of CPTP map divisibility and find that the results of steering scenario $\mathcal{G}^{An\to S}$ can have an agreement with the CPTP scheme.

Our work sheds light on understanding non-Markovian dynamics and quantum steering in open quantum systems, which also indicates that quantum steering has potential applications in diverse research domains, for instance, quantum synchronization~\cite{MariPRL2013,LiPRE2017,LiPRA2022}, quantum thermodynamics~\cite{CavinaPRL2017,RivasPRL2022,CollaPRA2022}, among others.

Furthermore, our research provides a robust and versatile framework for future experimental studies. The all-optical collision model scheme, with its inherent advantages of high precision, control, versatility, and efficient detection~\cite{ChiuriSR2012,CuevasSR2019}, offers a promising avenue for further exploration. We believe our findings will inspire and stimulate research interest in this important study area.

\section*{ACKNOWLEDGMENTS}
This work is supported by National Natural Science Foundation of China under Grant No. 11975064.

\begin{widetext}
\appendix
\section{COVARIANCE MATRICES}
\label{Sec:Appendix A}
This section aims to provide an overview of the distinct forms of the joint covariance matrices $\sigma_{J}$. As depicted in Eq.~(\ref{CovarianceMatrix}) in the main text, the detailed expression of $\sigma_{J}(j)$ is related to the scattering matrix of the model after finishing the $j$-th collision and demonstrates a specific pattern that can be obtained as follows:
\begin{equation}
V_{S} =
\begin{pmatrix}
\sigma_{x_1x_1} & \sigma_{x_1p_1} \\
\sigma_{x_1p_1} & \sigma_{p_1p_1}
\end{pmatrix},\quad V_{An} =
\begin{pmatrix}
\cosh \xi & 0 \\
0 & \cosh \xi
\end{pmatrix},\quad
V_{J} =
\begin{pmatrix}
\sinh \xi \text{Re}[C_{2,2}^{*}(j)] & \sinh \xi \text{Im}[C_{2,2}^{*}(j)]\\
\sinh \xi \text{Im}[C_{2,2}^{*}(j)] & -\sinh \xi \text{Re}[C_{2,2}^{*}(j)]
\end{pmatrix}.
\label{Eq:App_CMDetail}
\end{equation}
The detailed matrix elements of Eq.~(\ref{Eq:App_CMDetail}) are
\begin{equation}
\begin{aligned}
\sigma_{x_1x_1} = &(n + \frac{1}{2})  \sinh \zeta [e^{i\phi} C_{\text{m.e.}}^{'}+ e^{-i\phi} C_{\text{m.e.}}] + \cosh \xi |C_{2,2}(j)|^2 + (2n + 1)\cosh \zeta (1 - |C_{2,2}(j)|^2) ,\\
\sigma_{x_1p_1} = & \frac{1}{i}(n+\frac{1}{2}) \sinh \zeta [e^{i\phi} C_{\text{m.e.}}^{'} - e^{-i\phi} C_{\text{m.e.}}] ,\\
\sigma_{p_1p_1} = & -(n + \frac{1}{2}) \sinh \zeta [e^{i\phi} C_{\text{m.e.}}^{'} + e^{-i\phi} C_{\text{m.e.}}] + \cosh \xi |C_{2,2}(j)|^2 + (2n + 1)\cosh \zeta (1 - |C_{2,2}(j)|^2),
\end{aligned}
\end{equation}
where $C_{\text{m.e.}}=\sum_{l=3}^{L}{C_{l,2}(j)}^2$ and $C_{\text{m.e.}}^{'}=\sum_{l=3}^{L}{C_{l,2}^{*}(j)}^2$ denote the summations of matrix elements, and $C_{l,k}$ indicates the matrix element in the $l$-th row and the $k$-th column of scattering matrix $\mathcal{S}^{-1}(L)$, which is revealed in Eq.~(\ref{ChaFunINOUT}). The definitions of the other parameters have also been given in the main text.

\section{measure of non-Markovianity based on violation of divisibility}
\label{Sec:Appendix B}
In the main text, we have outlined the fundamental concepts and theoretical framework of the measure based on the violation of CPTP map divisibility, as discussed in Sec.~\ref{SEC:CPTP}. This appendix provides further elaboration on the measure's specifics.

As previously stated, the covariance matrix of the system after $j$-th collisions, denoted as $\sigma_S(j)$, can be represented as $\sigma_S(j) = \mathcal{E}_j[\sigma_S^{\text{in}}] = X_j\sigma_S^{\text{in}}X_j^{T} + Y_j$~\cite{TorrePRL2015}. This can be achieved by applying a mapping of $j$ collisions on the initial covariance matrix. To simplify the expression for $X_j$ and $Y_j$, we can express the corresponding elements as:

\begin{equation}
X_{j} =
\begin{pmatrix}
\text{Re}[C_{2,2}(j)] & -\text{Im}[C_{2,2}(j)]\\
\text{Im}[C_{2,2}(j)] & \text{Re}[C_{2,2}(j)]
\end{pmatrix},
\end{equation}
\begin{equation}
Y_{j} = N
\begin{pmatrix}
1 - \lvert C_{2,2}(j) \rvert^2 & 0\\
0 & 1 - \lvert C_{2,2}(j) \rvert^2
\end{pmatrix} + M \cos \phi
\begin{pmatrix}
    \text{Re}[ C_{\text{m.e.}}] & -\text{Im}[ C_{\text{m.e.}}]\\
    -\text{Im}[ C_{\text{m.e.}}] & -\text{Re}[  C_{\text{m.e.}}]
\end{pmatrix} + M \sin \phi
\begin{pmatrix}
    \text{Im}[  C_{\text{m.e.}}] & \text{Re}[  C_{\text{m.e.}}]\\
    \text{Re}[  C_{\text{m.e.}}] & -\text{Im}[  C_{\text{m.e.}}]
\end{pmatrix},
\end{equation}
where $N = (2n+1)\cosh \zeta$, $M=(2n+1)\sinh \zeta$, and the squeezing angle $\phi$ is corresponding to the environmental mode.

From this, we can give the explicit form of the Eq.~(\ref{Eq_juzhen}) as follows,
\begin{equation}
\mathcal{F}_{j} = (1 - \frac{\lvert C_{2,2}(j)\rvert^2}{ \lvert C_{2,2}(j - 1) \rvert^2})
\begin{pmatrix}
M\cos \phi + N & M\sin \phi - \frac{i}{2} \\
M\sin \phi + \frac{i}{2} & -M\cos\phi + N
\end{pmatrix}.
\end{equation}
The eigenvalues of $\mathcal{F}_{j}$ are related to the CPTP property of $\Phi_{j,j-1}$, and the expressions of the eigenvalues are
\begin{equation}
\nu_{j,\pm} = \frac{1}{2}(N \pm \sqrt{4M^2 + 1})(1 - \frac{\lvert C_{2,2}(j)\rvert ^2}{\lvert C_{2,2}(j - 1) \rvert ^2}).
\label{eigenvalue}
\end{equation}

If we consider the environment modes to be in a vacuum state, the eigenvalues of $\mathcal{F}{j}$ are $\nu_{j,+} = 0$ and $\nu_{j,-} = 1 - \frac{\lvert C_{2,2}(j) \rvert ^2}{\lvert C_{2,2}(j - 1)\rvert ^2}$~\cite{JinNJP2018}.

Consequently, we can calculate the non-Markovianity of Gaussian dissipation channel with a vacuum environment, denoted as $\mathcal{N}_{\text{CPTP,vac}}(L)$, using the following expression:
\begin{equation}
\mathcal{N}_{\text{CPTP,vac}}(L) = \sum_{j=2}^L \text{min}\lbrace0,1 - \frac{\lvert C_{2,2}(j)\rvert ^2}{\lvert C_{2,2}(j - 1)\rvert ^2}\rbrace.
\end{equation}
The above expression implies that the Gaussian dissipation channel with a vacuum environment exhibits non-Markovianity if and only if the following condition is satisfied:
\begin{equation}
|C_{2,2}(j)|^2 > |C_{2,2}(j - 1)|^2, \forall j \ge 2.
\end{equation}

Furthermore, if we consider the dissipation channel in the presence of a generic Gaussian environment, we can determine the non-Markovianity, $\mathcal{N}_{\text{CPTP, gau}}(L)$, by applying the following analysis~\cite{Li2022Entropy}:
\begin{equation}
\mathcal{N}_{\text{CPTP, gau}}(L) = (2n + 1)\cosh \zeta \mathcal{N}_{\text{CPTP,vac}}(L).
\end{equation}

\section{Gaussian steerability condition}
\label{Sec:Appendix C}
In this section, we present the Gaussian steerability conditions for $\mathcal{G}^{S \to An}$ and $\mathcal{G}^{An \to S}$, which involve the determinants of $V_{An}$, $V_S$, and $\sigma_J$. The specific expressions for these determinants are as follows:
\begin{equation}
\begin{aligned}
&\text{det}[V_{An}] = \cosh ^2 \xi,\\
&\text{det}[V_S] = [2(n + \frac{1}{2})(1 - \lvert C_{2,2}(j) \rvert^2)\cosh \zeta + \lvert C_{2,2}(j) \rvert^2 \cosh \xi ]^2 - [2(n+\frac{1}{2})(1 - \lvert C_{2,2}(j)\rvert^2)\sinh \zeta]^2,\\
&\text{det}[\sigma_{J}] = [\lvert C_{2,2}(j) \rvert^2 + 2(n + \frac{1}{2})(1 - \lvert C_{2,2}(j) \rvert^2)\cosh \zeta \cosh \xi]^2 - [2(n+\frac{1}{2})(1 - \lvert C_{2,2}(j) \rvert^2) \sinh \zeta \cosh \xi]^2,
\end{aligned}
\end{equation}
where we utilize the condition $\sum_{l=3}^{L} \lvert C_{l,2}(j) \rvert^2 = 1 - \lvert C_{2,2}(j) \rvert^2$  that is satisfied at any number of collisions $j$. From Eq.~(\ref{Eq:GaussianSteeringS}), we can conclude that in order to ensure that the system mode $S$ can steer the ancilla mode $An$, the determinant of $V_A$ must be greater than or equal to the determinant of $\sigma_{AB}$, which can be expressed as:
\begin{equation}
\begin{aligned}
\text{det}[V_S] - \text{det}[\sigma_{J}] &= [\lvert C_{2,2}(j) \rvert^4 - (2n+1)^2(1 - \lvert C_{2,2}(j) \rvert^2)^2 \cosh^2 \zeta](\cosh^2 \xi - 1) + [(2n+1)^2 (1 - \lvert C_{2,2}(j) \rvert^2)^2 \sinh^2 \zeta] (\cosh^2 \xi - 1)\\
&=(\cosh^2 \xi - 1) [\lvert C_{2,2}(j) \rvert ^4 - (2n + 1)^2(1 - \lvert C_{2,2}(j) \rvert ^2)^2] \ge 0
\end{aligned}
\end{equation}
Since $\cosh^2 \xi\ge 1$, we have $\lvert C_{2,2}(j) \rvert ^4 \ge (2n + 1)^2(1 - \lvert C_{2,2}(j) \rvert ^2)$, which must be satisfied for $\text{det}[V_A]$ to be greater than or equal to $\text{det}[\sigma_{AB}]$. Rearranging the inequality, we obtain the condition for the system mode $S$ to steer the ancilla mode $An$:
\begin{equation}
    \lvert C_{2,2}(j) \rvert^2 \ge 1 - \frac{1}{2(n+1)}.
\end{equation}

Next, we derive the conditions for $\mathcal{G}^{An \rightarrow S}$, which involves the inequality between $\text{det}[V_{An}]$ and $\text{det}[\sigma_{J}]$. The inequality can be given by:
\begin{equation}
 \text{det}[V_{An}] - \text{det}[\sigma_{J}]=\cosh^2 \xi - \lvert C_{2,2}(j) \rvert^4 - (2n+1)^2(1 - \lvert C_{2,2}(j) \rvert^2)^2 \cosh^2 \xi - 2(2n+1)(1 - \lvert C_{2,2}(j) \rvert^2)\lvert C_{2,2}(j) \rvert^2 \cosh \zeta \cosh \xi \ge 0
\label{Case2inequlityTotal}
\end{equation}

While obtaining a simple form for the above inequality is difficult, we are able to discuss it based on some specific environmental state settings. First, considering the case when the environmental modes are in thermal states ($\zeta = 0$), Eq.~(\ref{Case2inequlityTotal}) can be in this form
\begin{equation}
\cosh \xi - \lvert C_{2,2}(j) \rvert^2 - (2n+1)(1 - \lvert C_{2,2}(j) \rvert^2)\cosh \xi \ge 0
\end{equation}
Thus, we can obtain the inequality
\begin{equation}
\lvert C_{2,2}(j) \rvert^2 \ge \frac{2n\cosh\xi}{(2n+1)\cosh\xi-1}.
\label{eqCase2ThermalState}
\end{equation}
It can be found that if we determine the squeezing strength $\xi$ of the initial steerable entangled states $\rho_{J}$, we can have the detail conditions of  \textit{Gaussian An $\rightarrow$ S steerability} for different thermal environment states. In particular, when the environment modes to be vacuum state ($n=0$), Eq.~(\ref{eqCase2ThermalState}) reduces to $|C_{2,2}(j)|^2\ge0$, which will be constantly established.

Then, we consider the environment modes to be squeezed vacuum states, that is, $n=0$ and $\zeta \neq 0$, and the condition will be
\begin{equation}
(2\cosh \zeta \cosh \xi - \cosh^2 \xi - 1)\lvert C_{2,2}(j) \rvert^2 -2(\cosh \zeta \cosh \xi - \cosh^2 \xi)\ge 0.
\end{equation}
We will obtain the following relation:
\begin{equation}
\lvert C_{2,2}(j) \rvert^2 \ge \frac{2\cosh \xi (\cosh \xi - \cosh \zeta)}{1 - 2\cosh \zeta \cosh
\xi + \cosh^2 \xi}
\label{eqCase2Squeezd}
\end{equation}
The discussion above shows that the conditions for $\mathcal{G}^{An\rightarrow S}$ can also be derived by fixing $\xi$. Specifically, when $\xi = \zeta$, Eq.~(\ref{eqCase2Squeezd}) reduces to $|C_{2,2}(j)|^2\ge0$, which will always hold.
\end{widetext}


\begin{thebibliography}{9}
\bibitem{EPR1935PR}
A. Einstein, B. Podolsky, and N. Rosen, Can Quantum-Mechanical Description of Physical Reality Be Considered Complete?, Phys. Rev. {\bf47}, 777 (1935).

\bibitem{Cavalcanti2017RPP}
D. Cavalcanti, P. Skrzypczyk, Quantum steering: a review with focus on semidefinite programming, Rep. Prog. Phys. {\bf80} 024001 (2017).
\bibitem{Uola2020RMP}
R. Uola, A. C. S. Costa, H. C. Nguyen, and O. G\"uhne, Quantum steering, Rev. Mod. Phys. {\bf92}, 015001 (2020).
\bibitem{XiangPRXquantum2022}
Y. Xiang, S. Cheng, Q. Gong, Z. Ficek, and Q. He, Quantum Steering: Practical Challenges and Future Directions, PRX Quantum {\bf3}, 030102 (2022).

\bibitem{Wiseman2007PRL}
H. M. Wiseman, S. J. Jones, and A. C. Doherty, Steering, entanglement, nonlocality, and the Einstein-Podolsky-Rosen paradox, Phys. Rev. Lett. {\bf98}, 140402 (2007).

\bibitem{LiPRA2015}
C.-M. Li, Y.-N. Chen, N. Lambert, C.-Y. Chiu, and F. Nori, Certifying single-system steering for quantum-information processing, Phys. Rev. A {\bf92}, 062310 (2015).
\bibitem{XiangPRAR2017}
Y. Xiang, I. Kogias, G. Adesso, and Q. He, Multipartite Gaussian steering: Monogamy constraints and quantum cryptography applications, Phys. Rev. A {\bf95}, 010101(R) (2017).
\bibitem{HePRA2018}
G. P. He, Chaos in quantum steering in high-dimensional systems, Phys. Rev. A {\bf97}, 042340 (2018).

\bibitem{YadinNC2021}
B. Yadin, M. Fadel, and M. Gessner, Metrological complementarity reveals the Einstein-Podolsky-Rosen paradox, Nat. Commun. {\bf12}, 2410 (2021).
\bibitem{LeePRR2023}
K. -Y. Lee, J. -D. Lin, A. Miranowicz, F. Nori, H. -Y. Ku, and Y. -N. Chen, Steering-enhanced quantum metrology using superpositions of noisy phase shifts, Phys. Rev. Research {\bf5}, 013103 (2023).

\bibitem{HuangPRA2019}
C. -Y. Huang, N. Lambert, C. -M. Li, Y. -T. Lu, and F. Nori, Securing quantum networking tasks with multipartite Einstein-Podolsky-Rosen steering, Phys. Rev. A {\bf99}, 012302 (2019).
\bibitem{LuPRL2020}
H. Lu, C. -Y. Huang, Z. -D. Li, X. -F. Yin, R. Zhang, T. -L. Liao, Y. -A. Chen, C. -M. Li, and J. -W. Pan, Counting Classical Nodes in Quantum Networks, Phys. Rev. Lett. {\bf124}, 180503 (2020).
\bibitem{JonesPRL2021}
B. D. M. Jones, I. \v{S}upi\'{c}, R. Uola, N. Brunner, and P. Skrzypczyk, Network Quantum Steering, Phys. Rev. Lett. {\bf127}, 170405 (2021).

\bibitem{LinPRA2021}
J. -D. Lin, W. -Y. Lin, H. -Y. Ku, N. Lambert, Y. -N. Chen, and F. Nori, Quantum steering as a witness of quantum scrambling, Phys. Rev. A {\bf104}, 022614 (2021).
\bibitem{GheorghiuNJP2017}
A. Gheorghiu1, P. Wallden1, and E. Kashefi, Rigidity of quantum steering and one-sided device-independent verifiable quantum computation, New J. Phys. {\bf19} 023043 (2017).
\bibitem{Chen2017PRL}
S. -L. Chen, N. Lambert, C. -M. Li, A. Miranowicz, Y. -N. Chen, and F. Nori, Quantifying Non-Markovianity with Temporal Steering, Phys. Rev. Lett. {\bf 116}, 020503 (2016).
\bibitem{WangPRL2023}
Y. Wang, Z. -Y. Hao, J. -K. Li, Z. -H. Liu, K. Sun, J. -S. Xu, C. -F. Li, and G. -C. Guo, Observation of Non-Markovian Evolution of Einstein-Podolsky-Rosen Steering, Phys. Rev. Lett. {\bf130}, 200202 (2023).

\bibitem{Angel2014Quantum}
{\'A}. Rivas, S. F. Huelga, and M. B. Plenio, Quantum non-Markovianity: characterization, quantification, and detection, Rep. Prog. Phys. {\bf 77} 094001 (2014).
\bibitem{breuer2016colloquium}
H. -P. Breuer, E. -M. Laine, J. Piilo, and B. Vacchini, Colloquium: Non-Markovian dynamics in open quantum systems, Rev. Mod. Phys. {\bf88}, 021002 (2016).
\bibitem{VegaRMP2017}
I. d. Vega and D. Alonso, Dynamics of non-Markovian open quantum systems, Rev. Mod. Phys. {\bf89}, 015001 (2017).
\bibitem{AhnPRA2002}
D. Ahn, J. Lee, M. S. Kim, and S. W. Hwang, Self-consistent non-Markovian theory of a quantum-state evolution for quantum-information processing, Phys. Rev. A {\bf66}, 012302 (2002).
\bibitem{XuAoP2015}
Z. -Y. Xu, C. Liu, S. Luo b, and S. Zhu, Non-Markovian effect on remote state preparation, Annals of Physics {\bf356}, 29 (2015).
\bibitem{WhiteNC2020}
G. A. L. White, C. D. Hill, F. A. Pollock, L. C. L. Hollenberg, and K. Modi, Demonstration of non-Markovian process characterisation and control on a quantum processor. Nat. Commun. {\bf11}, 6301 (2020).

\bibitem{Crispin2004}
G. Crispin, and P. Zoller. {\it Quantum noise: a handbook of Markovian and non-Markovian quantum stochastic methods with applications to quantum optics}. (Springer Science \& Business Media, 2004).
\bibitem{TudelaQuantum2018}
A. Gonz\'{a}lez-Tudela, J. I. Cirac, Non-Markovian Quantum Optics with Three-Dimensional State-Dependent Optical Lattices, Quantum {\bf2}, 97 (2018).

\bibitem{WhitneyPRB2021}
R. S. Whitney, Non-Markovian quantum thermodynamics: Laws and fluctuation theorems, Phys. Rev. B {\bf98}, 085415 (2021).
\bibitem{DasPRA2021}
A. Das, A. Bera, S. Chakraborty, and D. Chru\'{s}ci\'{n}ski, Thermodynamics and the quantum speed limit in the non-Markovian regime, Phys. Rev. A {\bf104}, 042202 (2021).
\bibitem{CockrellPRE2022}
C. Cockrell and I. J. Ford, Stochastic thermodynamics in a non-Markovian dynamical system, Phys. Rev. E {\bf105}, 064124 (2022).
\bibitem{ChoquehuancaPRA2023}
J. M. Z. Choquehuanca, F. M. de Paula, and M. S. Sarandy, Non-Markovianity through entropy-based quantum thermodynamics, Phys. Rev. A {\bf107}, 012220 (2023).

\bibitem{BreuerPRL2009}
H.-P. Breuer, E.-M. Laine, and J. Piilo, Measure for the Degree of Non-Markovian Behavior of Quantum Processes in Open Systems, Phys. Rev. Lett. {\bf103}, 210401 (2009).
\bibitem{RivasPRL2010}
A. Rivas, S. F. Huelga, and M. B. Plenio, Entanglement and Non-Markovianity of Quantum Evolutions, Phys. Rev. Lett. {\bf105}, 050403 (2010).
\bibitem{TorrePRL2015}
G. Torre, W. Roga, and F. Illuminati, Non-Markovianity of Gaussian Channels, Phys. Rev. Lett. {\bf115}, 070401 (2015).
\bibitem{TorrePRA2018}
G. Torre and F. Illuminati, Exact non-Markovian dynamics of Gaussian quantum channels: Finite-time and asymptotic regimes, Phys. Rev. A {\bf98}, 012124 (2018).

\bibitem{HolevoJMP2011}
A. S. Holevo, The Choi-Jamiolkowski forms of quantum Gaussian channels, J. Math. Phys. (N.Y.) {\bf 52}, 042202 (2011).
\bibitem{VasilePRA2011}
R. Vasile, S. Maniscalco, M. G. A. Paris, H. -P. Breuer, and J. Piilo, Quantifying non-Markovianity of continuous-variable Gaussian dynamical maps, Phys. Rev. A {\bf 84}, 052118 (2011).

\bibitem{BreuerBook2002}
F. Petruccione and H. P. Breuer, {\it The theory of open quantum systems}(Oxford University, New York, 2002).
\bibitem{GiovannettiPRL2021}
V. Giovannetti and G. M. Palma, Master Equations for Correlated Quantum Channels, Phys. Rev. Lett. {\bf108}, 040401 (2012).

\bibitem{CiccarelloPRA2013}
F. Ciccarello, G. M. Palma, and V. Giovannetti, Collision-model-based approach to non-Markovian quantum dynamics, Phys. Rev. A {\bf87}, 040103(R) (2013).
\bibitem{JinPRA2015}
J. Jin, V. Giovannetti, R. Fazio, F. Sciarrino, P. Mataloni, A. Crespi, and R. Osellame, All-optical non-Markovian stroboscopic quantum simulator, Phys. Rev. A {\bf91}, 012122 (2015).
\bibitem{CattaneoPRL2021}
M. Cattaneo, G. D. Chiara, S. Maniscalco, R. Zambrini, and G. L. Giorgi, Collision Models Can Efficiently Simulate Any Multipartite Markovian Quantum Dynamics, Phys. Rev. Lett. {\bf126}, 130403 (2021).
\bibitem{CiccarelloPR2022}
F. Ciccarello, S. Lorenzo, V. Giovannetti, G. M. Palma, Quantum collision models: Open system dynamics from repeated interactions, Phys. Rep. {\bf954}, 1 (2022).

\bibitem{CamascaPRA2021}
R. R. Camasca and G. T. Landi, Memory kernel and divisibility of Gaussian collisional models, Phys. Rev. A {\bf103}, 022202 (2021).
\bibitem{Li2022Entropy}
Y. Li, X. Li, and J. Jin, Dissipation-Induced Information Scrambling in a Collision Model, Entropy {\bf24}, 345 (2022).
\bibitem{ZhangPRA2021}
Q. Zhang, Z. -X. Man, and Y. -J. Xia, Non-Markovianity and the Landauer principle in composite thermal environments, Phys. Rev. A {\bf103}, 032201 (2021).
\bibitem{ZhangPRA2023v1}
Q. Zhang, Y. -J. Xia, and Z. -X. Man, Effects of one-way correlations on thermodynamics of a multipartite open quantum system, Phys. Rev. A {\bf108}, 062211 (2023).
\bibitem{ZhangPRA2023v2}
Q. Zhang, Z. -X. Man, Y. -J. Zhang, W. -B. Yan, and Y. -J. Xia, Quantum thermodynamics in nonequilibrium reservoirs: Landauer-like bound and its implications, Phys. Rev. A {\bf107}, 042202 (2023).
\bibitem{LiPRA2023}
Y. Li, X. Li, and J. Jin, Quantum nonstationary phenomena of spin systems in collision models, Phys. Rev. A {\bf107}, 042205 (2023).

\bibitem{KarpatPRA2019}
G. Karpat, \.{I}. Yal\c{c}\textsc{i}nkaya, and B. \c{C}akmak, Quantum synchronization in a collision model, Phys. Rev. A {\bf100}, 012133 (2019).
\bibitem{KarpatPRA2020}
G. Karpat, \.{I}. Yal\c{c}\textsc{i}nkaya, and B. \c{C}akmak, Quantum synchronization of few-body systems under collective dissipation, Phys. Rev. A {\bf101}, 042121 (2020).

\bibitem{ArisoyEntropy2019}
O. Ar\textsc{i}soy, S. Campbell, \"{O}. E. M\"{u}stecapl\textsc{i}o\v{g}lu, Thermalization of Finite Many-Body Systems by a Collision Model, Entropy, {\bf21}, 1182 (2019).
\bibitem{LiPRA2020}
Y. Li, X. Li, and J. Jin, Information scrambling in a collision model, Phys. Rev. A {\bf101}, 042324 (2020).

\bibitem{FrigerioPRA2021}
M. Frigerio, S. Hesabi, D. Afshar, and M. G. A. Paris, Exploiting Gaussian steering to probe non-Markovianity due to the interaction with a structured environment, Phys. Rev. A {\bf104}, 052203 (2021).
\bibitem{SantisNJP2023}
D. D. Santis, D. Farina, M. Mehboudi, and A. Ac\'{i}n, Ancillary Gaussian modes activate the potential to witness non-Markovianity, New J. Phys. {\bf25} 023025 (2023).
\bibitem{KyawPRA2020}
T. H. Kyaw, V. M. Bastidas, J. Tangpanitanon, G. Romero, and L. -C. Kwek, Dynamical quantum phase transitions and non-Markovian dynamics, Phys. Rev. A {\bf101}, 012111 (2020).

\bibitem{ShibataJSP1977}
F. Shibata, Y. Takahashi, and N. Hashitsume, A generalized stochastic liouville equation. Non-Markovian versus memoryless master equations, J. Stat. Phys. {\bf17}, 171 (1977).
\bibitem{ChaturvediZPB1979}
S. Chaturvedi and F. Shibata, Time-convolutionless projection operator formalism for elimination of fast variables. Applications to Brownian motion, Z. Phys. B {\bf35}, 297 (1979).
\bibitem{UchiyamaPRE1999}
C. Uchiyama and F. Shibata, Unified projection operator formalism in nonequilibrium statistical mechanics, Phys. Rev. E {\bf60}, 2636 (1999).
\bibitem{BreuerAof2001}
H.-P. Breuer, B. Kappler, and F. Petruccione, The Time-Convolutionless Projection Operator Technique in the Quantum Theory of Dissipation and Decoherence, Annals of Physics {\bf291}, 36 (2001).

\bibitem{NakajimaPTP1958}
S. Nakajima, On Quantum Theory of Transport Phenomena: Steady Diffusion, Prog. Theor. Phys. {\bf20}, 948 (1958).
\bibitem{ZwanzigJCP1960}
R. Zwanzig, Ensemble Method in the Theory of Irreversibility, J. Chem. Phys. {\bf33}, 1338 (1960).
\bibitem{VacchiniPRA2010}
B. Vacchini and H. -P. Breuer, Exact master equations for the non-Markovian decay of a qubit, Phys. Rev. A {\bf81}, 042103 (2010).
\bibitem{IvanovPRA2015}
A. Ivanov and H. -P. Breuer, Extension of the Nakajima-Zwanzig approach to multitime correlation functions of open systems, Phys. Rev. A {\bf92}, 032113 (2015).

\bibitem{DengPRL2023}
Y. -H. Deng, S. -Q. Gong, Y. -C. Gu, Z. -J. Zhang, H. -L. Liu, H. Su, H. -Y. Tang, J. -M. Xu, M. -H. Jia, M. -C. Chen, H. -S. Zhong, H. W., J. Yan, Y. Hu, J. Huang, W. -J. Zhang, H. Li, X. Jiang, L. You, Z. Wang, L. Li, N. -L. Liu, C. -Y. Lu, and J. -W. Pan, Solving Graph Problems Using Gaussian Boson Sampling, Phys. Rev. Lett. {\bf130}, 190601 (2023).

\bibitem{ZhongPRL2021}
H. -S. Zhong {\it et al}., Phase-Programmable Gaussian Boson Sampling Using Stimulated Squeezed Light, Phys. Rev. Lett. {\bf127}, 180502 (2021).

\bibitem{ChiuriSR2012}
A. Chiuri, C. Greganti, L. Mazzola, M. Paternostro, and P. Mataloni , Linear Optics Simulation of Quantum Non-Markovian Dynamics, Sci. Rep. {\bf2}, 968 (2012).
\bibitem{CuevasSR2019}
\'{A}. Cuevas, A. Geraldi, C. Liorni, L. D. Bonavena, A. D. Pasquale, F. Sciarrino, V. Giovannetti, and P. Mataloni, All-optical implementation of collision-based evolutions of open quantum systems, Sci. Rep. {\bf9}, 3205 (2019).

\bibitem{SouzaPRA2015}
L. A. M. Souza, H. S. Dhar, M. N. Bera, P. Liuzzo-Scorpo, and G. Adesso, Gaussian interferometric power as a measure of continuous-variable non-Markovianity, Phys. Rev. A {\bf92}, 052122 (2015).

\bibitem{WallsQO1994}
D. F. Walls and G. J. Milburn, {\it Quantum Optics} (Springer, Berlin, 1994).
\bibitem{JinNJP2018}
J. Jin and C.-s. Yu, Non-Markovianity in the collision model with environmental block, New J. Phys. {\bf20}, 053026 (2018).
\bibitem{WangPR2007}
X. -B. Wang, T. Hiroshima, A. Tomita, M. Hayashi, Quantum information with Gaussian states, Phys. Rep., {\bf448}, 1(2007).
\bibitem{Kogias2015PRL}
I. Kogias, A. R. Lee, S. Ragy, and G. Adesso, Quantification of Gaussian Quantum Steering, Phys. Rev. Lett. {\bf 114}, 060403 (2015).

\bibitem{footnote1}
The significant difference between the maximum and minimum values of the non-Markovianity degrees makes accurately distinguishing the boundary between the non-Markovian and Markovian regions based on the original data unfeasible. To address this issue, we have implemented a scaling method for the values of the measure in all heat maps.

\bibitem{HuEPJC2021}
B. Hu, C. Wen, J. Wang and J. Jing, Gaussian quantum steering under the influence of a dilaton black hole, Eur. Phys. J. C {\bf 81}, 925 (2021).
\bibitem{DengNPJQI2021}
X. Deng, Y. Liu, M. Wang, X. Su and K. Peng, Sudden death and revival of Gaussian Einstein-Podolsky-Rosen steering in noisy channels, npj Quantum Inf. {\bf7}, 65 (2021).
\bibitem{RosarioPLA2022}
P. Rosario, A. F. Ducuara and C. E. Susa, Quantum steering and quantum discord under noisy channels and entanglement swapping, Phys. Lett. A, {\bf 440}, 128144 (2022).
\bibitem{LiSR2023}
W.-C. Li, Y. Xiao, X.-H. Han, X. Fan, X.-B. Hei and Y. -J. Gu, Dynamics of multipartite quantum steering for different types of decoherence channels, Sci. Rep. {\bf13}, 3798 (2023).

\bibitem{MariPRL2013}
A. Mari, A. Farace, N. Didier, V. Giovannetti, and R. Fazio, Measures of Quantum Synchronization in Continuous Variable Systems, Phys. Rev. Lett. {\bf111}, 103605 (2013).
\bibitem{LiPRE2017}
W. Li, W. Zhang, C. Li, and H. Song, Properties and relative measure for quantifying quantum synchronization, Phys. Rev. E {\bf96}, 012211 (2017).
\bibitem{LiPRA2022}
W. Li, Analyzing quantum synchronization through Bohmian trajectories, Phys. Rev. A {\bf106}, 023512 (2022).

\bibitem{CavinaPRL2017}
V. Cavina, A. Mari, and V. Giovannetti, Slow Dynamics and Thermodynamics of Open Quantum Systems, Phys. Rev. Lett. {\bf119}, 050601 (2017).
\bibitem{RivasPRL2022}
\'{A}. Rivas, Strong Coupling Thermodynamics of Open Quantum Systems, Phys. Rev. Lett. {\bf124}, 160601 (2022).
\bibitem{CollaPRA2022}
A. Colla and H. -P. Breuer, Open-system approach to nonequilibrium quantum thermodynamics at arbitrary coupling, Phys. Rev. A {\bf105}, 052216 (2022).


\end{thebibliography}
\end{document}